\documentclass[10pt]{article}
\usepackage{amsmath}
\usepackage{graphicx}
\usepackage{color}
\usepackage{orcidlink}
\usepackage{hyperref}
\usepackage{multirow}
\usepackage{rotating}
\usepackage{multicol}
\hypersetup{colorlinks=true, linkcolor=blue, citecolor=blue, urlcolor=blue}
\usepackage{caption}
\usepackage{subcaption}
\usepackage{setspace}
\RequirePackage[numbers,sort&compress]{natbib}
\begin{document}
\baselineskip=14pt

\begin{center}
\LARGE{Gravitational Lensing and Topological Photon Sphere of Holonomy Corrected Schwarzschild Black Hole with a Cloud of Strings}
\par\end{center}

\vspace{0.3cm}

\begin{center}
{\bf Faizuddin Ahmed\orcidlink{0000-0003-2196-9622}}\footnote{\bf faizuddinahmed15@gmail.com}\\
{\it Department of Physics, Royal Global University, Guwahati, 781035, Assam, India}\\
\vspace{0.15cm}
{\bf Shubham Kala\orcidlink{0000-0003-2379-0204}}\footnote{\bf shubhamkala871@gmail.com (Corresp. author)}\\
{\it The Institute of Mathematical Sciences, C.I.T. Campus, Taramani, Chennai, 600113, Tamil Nadu, India}
\end{center}

\vspace{0.3cm}

\begin{abstract}
In this paper, we theoretically investigate the deflection of light, lensing equations, topological properties of photon rings, and accretion disk characteristics in the spacetime of a holonomy-corrected Schwarzschild black hole surrounded by a cloud of strings. The analysis is carried out in the weak-field limit, where we analytically derive expressions for the deflection angle and extract the corresponding lensing observables. These results reveal the dependence of light deflection on the string cloud parameter and the holonomy correction parameter, offering potential observational signatures of underlying quantum gravity effects. We model possible gravitational scenarios to explore the distinguishing features of this modified BH geometry and assess its deviation from classical solutions through gravitational lensing behavior. Furthermore, we analyze the topological structure of the photon sphere by constructing a normalized vector field and demonstrate how it is affected by the presence of string clouds and holonomy corrections. Finally, we examine the properties of a thin accretion disk in this BH background, showing that both the string cloud and holonomy parameters significantly influence the disk’s radiation profile, temperature distribution, and spectral characteristics. Our results suggest that these modifications leave measurable imprints, providing viable avenues for observational constraints in the strong-gravity regime.
\end{abstract}

\section{Introduction}

Despite the great scientific and technological advances provided by General Relativity (GR) \cite{tst}, this theory has problems with geodesic singularities, as is the case with black holes and the big bang \cite{1457,529}. Faced with this scenario, among other cosmological issues \cite{SSTC, SCPC, Knop, BoomerangCollaboration, Steinhardt, Halverson}, physicists have been working on alternative gravitational theories that are consistent with current observations and that are capable of avoiding geodesic singularities. Among these theories, we highlight Loop Quantum Gravity (LQG) \cite{Rovelli003, Ashtekar84, Perez80}. LQG is a non-perturbative theory for quantizing the structure of spacetime and, although it does not yet present a complete quantum description close to a singularity, it has presented effective models in low-energy regimes with corrections arising from quantum effects. Recently,
in \cite{Alonso-Bardaji829, Alonso-Bardaji106}, using LQG, the authors derived a spacetime solution corresponding to a singularity-free interior (black hole/white hole) and two asymptotically flat outer regions. The inner region contains a black-bounce surface, replacing the standard Schwarzschild spacetime singularity.

Gravitational lensing arises when a massive object or distribution of matter-such as a galaxy cluster-generates a gravitational field that significantly curves the surrounding space-time. This curvature alters the trajectory of light from distant sources, causing it to bend as it travels toward the observer. The deflection of light is due to the gravitational field’s influence on photon paths, leading to observable shifts in the apparent position and properties of the source as seen from Earth. This phenomenon plays a critical role in cosmology and gravitation, manifesting in both weak lensing (where light passes far from the lensing mass) and strong lensing (where light travels near massive objects). Gravitational lensing has been extensively studied across a wide range of space-time backgrounds, including charged BH in string theory~\cite{ref35a}, Schwarzschild BH~\cite{AA8,ref44a,AA10}, brane-world BHs~\cite{ref38a}, Reissner-Nordström BHs~\cite{ref39a}, space-times with naked singularities~\cite{ref40a, ref42a, ref43a,ref61a,ref62a,ref63a,ref64a}, Kerr-Randers optical geometries~\cite{ref45a}, and Eddington-inspired Born-Infeld (EiBI) theory with global monopoles~\cite{ref46a,ref57a} and cosmic strings~\cite{ref59a}. Further investigations include Kerr-MOG BHs~\cite{ref47a}, Simpson-Visser black-bounce space-times~\cite{ref48a}, rotating regular BHs~\cite{ref49a}, non-rotating black-bounce space-times~\cite{ref50a}, Rindler-modified Schwarzschild BHs~\cite{ref52a}, traversable wormhole~\cite{ref54a, ref55a}, topologically charged Ellis-Bronnikov wormhole~\cite{ref56a} and Morris-Thorne wormhole with cosmic strings~\cite{ref58a}, holonomy corrected Schwarzschild BH \cite{HH3,HH4}, topologically charged holonomy corrected Schwarzschild BH \cite{ref61aa}, and holonomy corrected Schwrazschild BH with phantom global monopoles \cite{ref60a}.

Gravitational lensing by BHs surrounded by exotic matter distributions such as string clouds and quintessence has attracted significant attention in recent years. Sharif \textit{et al.} investigated the weak lensing behavior of Schwarzschild BHs in the presence of a string cloud, showing how string cloud alter the deflection angle~\cite{GM1}. Similarly, Li and Zhou examined the weak gravitational lensing of a Kerr BH in Rastall gravity with a surrounding string cloud~\cite{GM4}, while Mustafa \textit{et al.} explored the shadows and deflection angles for various string cloud spacetimes under quintessence backgrounds~\cite{GM5}. More recent developments by Atamurotov \textit{et al.} include studies on gravitational lensing and shadows of rotating BHs in non-Kerr geometries and modified theories, particularly within symmetric frameworks~\cite{GM7}. Extensions to quantum gravity frameworks, including corrections from generalized uncertainty principles and higher-order curvature terms, have been discussed in the context of null geodesics and thermodynamic properties~\cite{GM2}. Incorporating rotation and charge, Zhao \textit{et al.} studied gravitational lensing in rotating regular BHs with string clouds, focusing on the deflection angle and shadow radius~\cite{GM3}. Molla \textit{et al.} presented a comprehensive analysis of the deflection angle and photon orbits for BHs influenced by string clouds, quintessence, and plasma effects~\cite{GM8}, while Atamurotov \textit{et al.} investigated BH shadows and geodesics in higher-dimensional and deformed spacetimes~\cite{GM9,GM10}. Other studies explored photon spheres and null trajectories in regular, rotating, and magnetically charged spacetimes~\cite{GM11,GM12}. The influence of topological defects, spin effects, and generalized optical metrics on lensing observables has also been studied in the context of brane-world BH and other exotic geometries~\cite{GM13,GM14}. A detailed numerical analysis of the effective potential and deflection angle in such extended geometries was recently conducted in~\cite{GM15}. Some recent studies on GL under weak and strong field limit in various geometric backgrounds were reported in \cite{Kala:2020prt,Kala:2020viz,Kala:2021ppi,Kala:2022uog,Vishvakarma:2024icz,Kala:2024fvg,Pantig:2024lpg,Kala:2025xnb,Roy:2025hdw,Kukreti:2025rzn,Kala:2025fld}.

To determine the existence and properties of photon spheres, several approaches can be employed, such as analyzing the effective potential, employing Hamiltonian formalism, utilizing Killing symmetries, or applying topological methods. In this work, we adopt the topological method to investigate the photon sphere structure of black holes. Cardoso et al.~\cite{ref32} argued that the presence of light rings (photon spheres) could serve as observational evidence for event horizons. They emphasized that the nonlinear instability of ultra-compact objects with light rings provides strong support for the black hole hypothesis, especially when such features are detected via electromagnetic or gravitational-wave observations. Cunha and collaborators further explored this concept by demonstrating that standard, spherically symmetric black holes admit circular photon orbits on a plane. Importantly, they showed that while unstable photon orbits can be linked to the observable shadows of black holes, stable ones can induce spacetime instabilities~\cite{ref33}. Building on this, they formulated a theorem stating that: Axisymmetric, stationary solutions of the Einstein field equations resulting from classical gravitational collapse of matter obeying the null energy condition, which are everywhere smooth and ultra-compact (i.e., possess a light ring), must contain at least two light rings-one of which is stable. The existence of stable light rings, as they argued, is typically associated with nonlinear instabilities in spacetime~\cite{ref34}.

This line of investigation was extended by Cunha and Herdeiro~\cite{ref36}, who tested the above hypotheses in the context of four-dimensional black hole solutions. They showed that: A stationary, axisymmetric, and asymptotically flat black hole spacetime in (3+1) dimensions, with a non-extremal, topologically spherical, and Killing horizon, admits at least one standard light ring outside the event horizon for each sense of rotation. Building upon these developments, Wei~\cite{ref37} employed the Duan topological current method to explore the global properties of photon spheres, demonstrating that at least one standard photon sphere exists outside black holes not only in asymptotically flat spacetimes but also in asymptotically AdS and dS geometries.  In a complementary study, Ghosh and Sarkar~\cite{ref35} also examined the nature of light rings in stationary spacetimes, further supporting the idea that their presence plays a fundamental role in characterizing black hole geometries. Following these foundational works, Sadeghi et al.~\cite{ref38} applied the topological method to a range of black hole models, including Einstein--Yang--Mills non-minimal black holes, AdS black holes surrounded by Chaplygin-like dark fluids, and Bardeen-like black holes in Einstein-Gauss-Bonnet gravity. Their analysis focused on identifying topological photon and anti-photon spheres, emphasizing the essential role these structures play in defining the physical characteristics of ultra-compact objects. They proposed a classification scheme for the parameter space of black hole solutions based on the presence and location of photon spheres. Building on this idea, Sadeghi et al.~\cite{ref39} extended the topological charge method to study photon spheres in black holes with hyperscaling violation (HSV), applying similar techniques as those developed earlier. Additionally, Liu et al.~\cite{ref40} implemented the same approach for two classes of static black hole solutions that preserve general covariance. Collectively, these studies offer compelling evidence that photon spheres and light rings are fundamental structural features of black holes. Motivated by this insight, we aim to explore whether the existence of photon spheres can be used to constrain the allowed ranges of specific parameters in black hole models. To achieve this, we apply the classification technique proposed by Cunha and Wei, based on the total topological charge associated with photon spheres.

In this work, we aim to investigate the behavior of null geodesics and the corresponding gravitational lensing effects in a class of spherically symmetric BHs surrounded by a cloud of strings, incorporating holonomy corrections motivated by loop quantum gravity frameworks. Our primary objective is to understand how these quantum corrections-stemming from the polymerization of the gravitational phase space-influence the propagation of photon trajectories and modify the associated gravitational lensing observables. We find that the deflection angle of photon rays in the weak field limit is significantly affected by both holonomy corrections and the presence of the string cloud. These modifications result in observable deviations from the classical predictions derived for the standard Schwarzschild BH, thereby highlighting the non-trivial influence of quantum geometric and topological structures. Beyond gravitational lensing, we derive the corresponding lens equations, which describe the bending of light rays in this background, and explicitly demonstrate how key geometric and physical parameters-such as the polymerization constant and the density of the string cloud-affect observable quantities like the image positions and magnifications. Furthermore, we also explore the topological features of the photon sphere in the chosen BH. Specifically, we analyze how the normalized vector field on the photon sphere is altered due to the combined effects of holonomy corrections and string cloud contributions, providing insights into the interplay between quantum gravity and topological characteristics of null trajectories. We demonstarte a clear differences of this vector field with the holonomy-corrected Schwarzschild BH and Letelier BH solutions. Finally, we investigate the properties of a geometrically thin and optically thick accretion disk surrounding the BH. We examine how the aforementioned corrections influence the disk's energy flux, temperature profile, and emission spectra. Our results indicate that holonomy effects and the presence of string clouds introduce measurable deviations in the curvature of the spacetime, thereby affecting the accretion dynamics and the electromagnetic signatures of the black hole environment.

This paper is organized as follows: In Section~\ref{sec:2}, we introduce the background BH geometry and investigate the behavior of null geodesics, along with the corresponding gravitational lensing phenomena. In Section~\ref{sec:3}, we derive the lens equation for the BH and analyze how the physical and geometric parameters influence the lensing observables. In Section~\ref{sec:4}, we explore the topological properties of the photon sphere and examine the effects of holonomy corrections and the string cloud on its structure. In Section~\ref{sec:5}, we discuss the dynamics and physical properties of a thin accretion disk surrounding the BH, focusing on how the aforementioned corrections affect its emission characteristics. Finally, in Section~\ref{sec:6}, we summarize our findings and present the main conclusions of the study.

\section{Holonomy-Corrected BH with a Cloud of Strings}

In this section, we introduce the background BH geometry that serves as the foundational setup for the study of optical and topological properties in the modified gravitational field. 

The metric line element that describes the holonomy corrected Schwarzschild BH in spherical coordinates ($t,r,\theta, \phi$), is given by \cite {HH1,HH2,HH3,HH4}
\begin{eqnarray}\label{aa}
	ds^2=-f(r)\,dt^2+\frac{dr^2}{f(r)\,\left(1-\ell/r\right)}+ r^2\,\left(d\theta^2+\sin^2\theta\, d\phi^2\right),\quad f(r)=1-\frac{2M}{r},
\end{eqnarray}
where \( M \) is the mass of BH, \( \ell \) is a new scale length defined by \(\ell = \frac{2\,m\, \lambda^2}{1 + \lambda^2}\) also called the LQG parameter, where \( \lambda \) is called the polymerization constant and provides the holonomy correction information. For \( m > 0 \), this solution is asymptotically flat and contains a globally hyperbolic BH or white hole region with a minimal space-like hypersurface replacing the original singularity.

The first studies concerning a formalism to treat gravity with a cloud of strings as source, in the framework of General Relativity (GR), were presented by Letelier~\cite{Letelier1979}. Using this formalism, he obtained a generalization of the Schwarzschild solution corresponding to a black hole surrounded by a spherically symmetric cloud of strings, whose energy-momentum tensor is given by
\begin{align}
T^t_{\ t}= T^r_{\ r} =\rho_c= \frac{\alpha}{r^2}, \quad T^\theta_{\ \theta} &= T^\phi_{\ \phi} = 0, \label{aa1}
\end{align}
where $\rho_c$ is the energy density of the cloud and \( \alpha \) is a constant associated with the presence of the string cloud. Solving the Einstein’s equations taking into account the source given by Eq. (\ref{aa1}), the metric of the Schwarzschild BH with string cloud context is given by
\begin{eqnarray}\label{aa2}
	ds^2=-F(r)\,dt^2+dr^2/F(r)+ r^2\,\left(d\theta^2+\sin^2\theta\, d\phi^2\right),\quad F(r)=1-\alpha-2M/r.
\end{eqnarray}

Motivated by the aforementioned studies, as well as the recent presented in~\cite{ref61aa}, we consider a class of static and spherically symmetric BH incorporating holonomy corrections and surrounded by a cloud of strings. The resulting space-time is described by the following line element:
\begin{eqnarray}\label{aa3}
	ds^2=-F(r)\,dt^2+\frac{dr^2}{F(r)\,\left(1-\ell/r\right)}+ r^2\,\left(d\theta^2+\sin^2\theta\, d\phi^2\right),
\end{eqnarray}
where symbols have their usual meanings as stated above. The parameters are in the ranges: $\alpha \in [0, 1)$ and $\ell \geq 0$. The ranges of the coordinates are as follows:
\begin{align}
& t \in (-\infty, \infty)\, && \text{(temporal coordinate)} \nonumber\\
& r \in (0, \infty) \,&& \text{(radial coordinate)} \nonumber\\
& \theta \in [0, \pi]\, && \text{(polar angle)} \nonumber\\
& \phi \in [0, 2\pi)\, && \text{(azimuthal angle)}.\label{range}
\end{align}

For this space-time (\ref{aa3}), we determine the Kretschmann scalar $\mathcal{K}=R^{\mu\nu\rho\sigma}\,R_{\mu\nu\rho\sigma}$, the quadratic Ricci tensor $R^{\mu\nu}\,R_{\mu\nu}$, and the Ricci scalar $R=g^{\mu\nu}\,R_{\mu\nu}$ and analyze the effects of string cloud and holonomy correction parameters.
\begin{itemize}
    \item The Krestchmann scalar for the given space-time is as follows:
    \begin{equation}
        \mathcal{K}=\frac{
3 M^2 \left(16 r^2 - 40 r \ell + 27 \ell^2 \right)
+ 8 M r \left[ r \ell (3 - 5 \alpha)
+ 4 \ell^2 ( -1 + \alpha)
+ 2 r^2 \alpha \right]
+ 2 r^2 \left[
3 \ell^2 ( -1 + \alpha)^2
- 4 r \ell ( -1 + \alpha) \alpha+ 2 r^2 \alpha^2\right]
}{r^8}.\label{aa4}
    \end{equation}

\item The quadratic Ricci tesnor is given by

\begin{equation}
    R^{\mu\nu}\,R_{\mu\nu}=\frac{
9 M^2 \ell^2 
+ 2 M r \ell \left[ \ell (-1 + \alpha) + 4 r \alpha \right]
+ r^2 \left[
3 \ell^2 (-1 + \alpha)^2 
- 4 r \ell (-1 + \alpha) \alpha 
+ 4 r^2 \alpha^2
\right]
}{2 r^8}.\label{aa5}
\end{equation}

    \item The Ricci scalar is given by

    \begin{equation}
        R=\frac{3 M \ell}{r^4}+\frac{2 \alpha}{r^2}.\label{aa6}
    \end{equation}
\end{itemize}

\begin{figure}[ht!]
    \centering
    \includegraphics[width=0.32\linewidth]{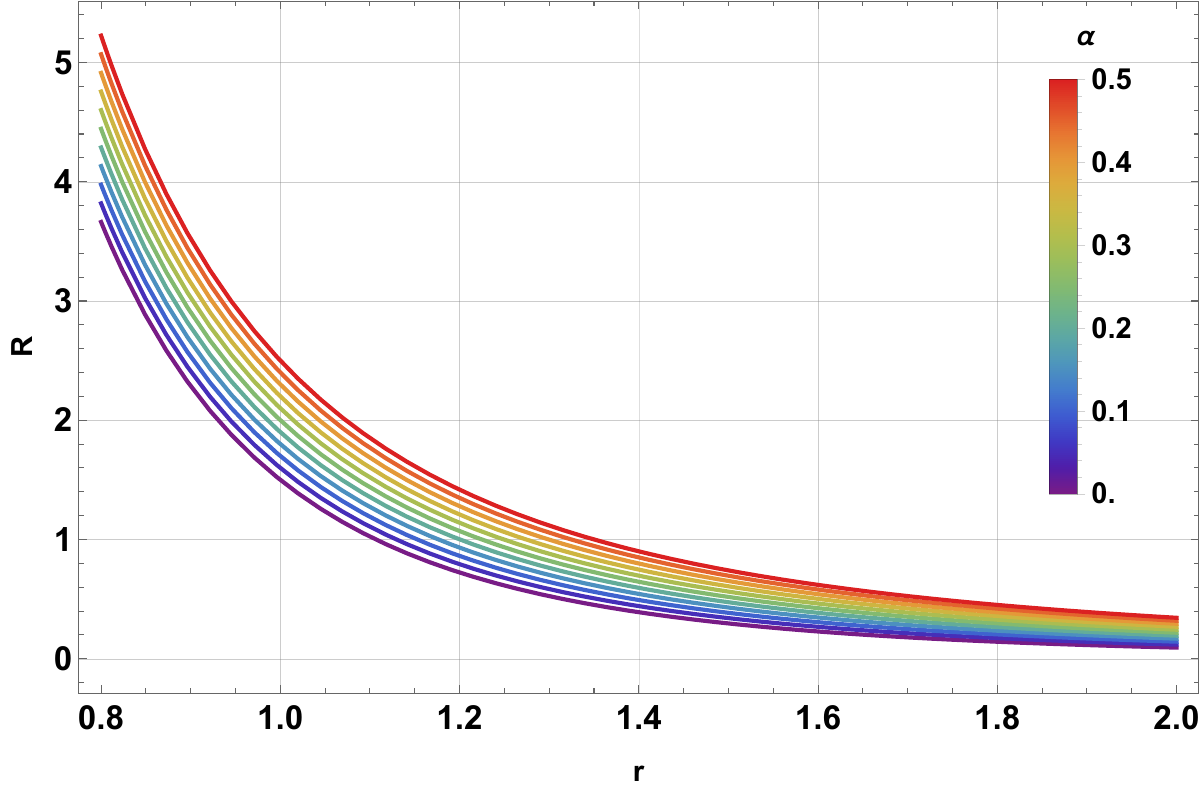}\quad
    \includegraphics[width=0.32\linewidth]{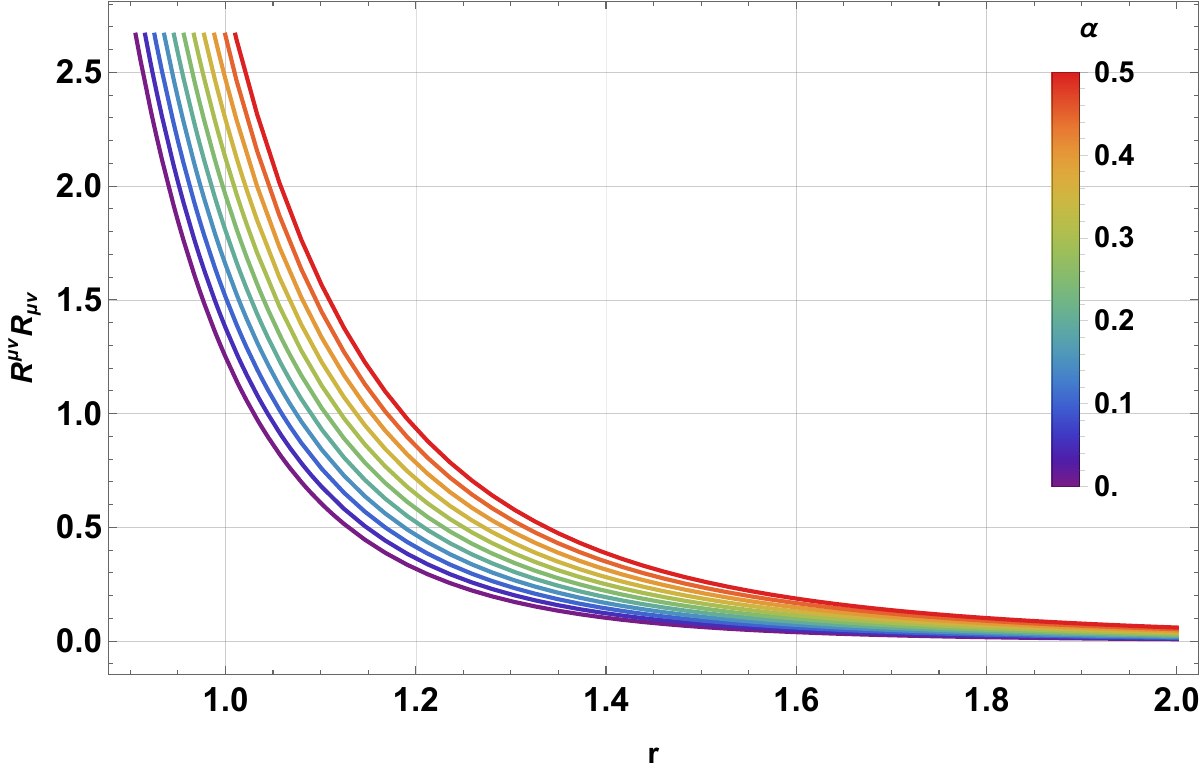}\quad
    \includegraphics[width=0.32\linewidth]{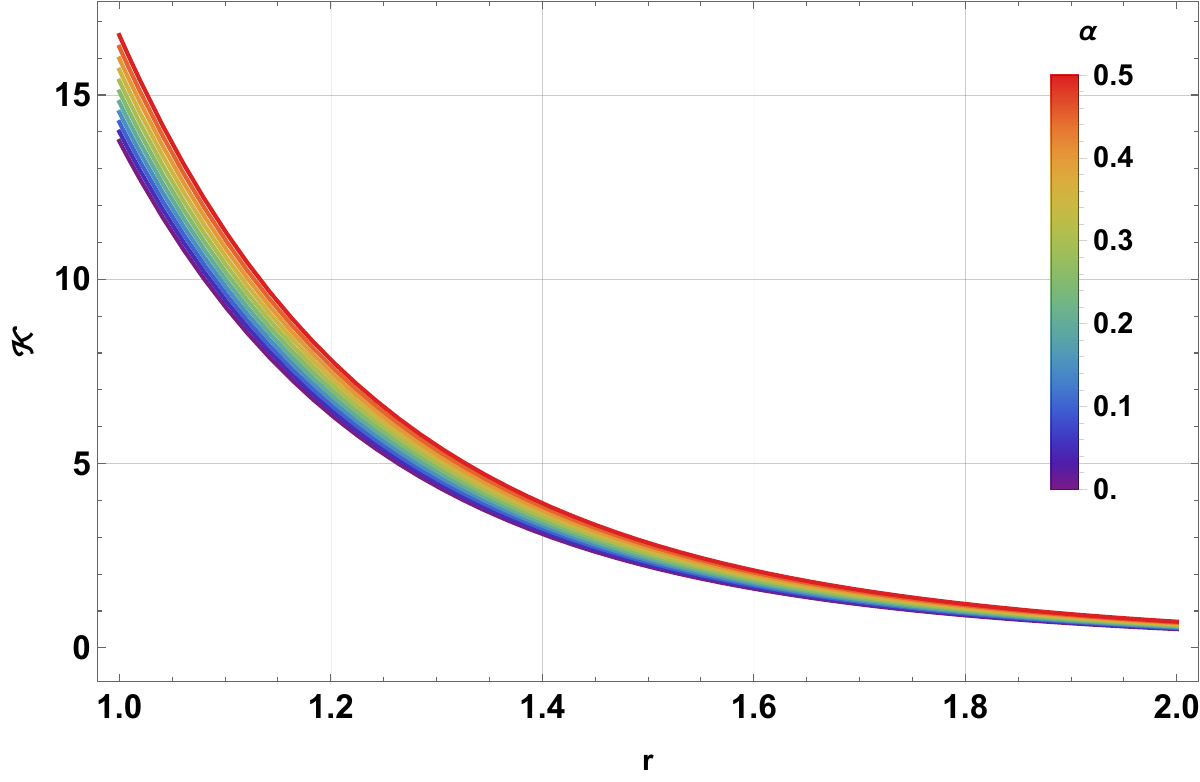}
    \caption{\footnotesize The behavior of various scalar curvatures for different values of CoS parameter $\alpha$, while the BH mass $M=1$ and holonomy-correction $\ell=0.5$.}
    \label{fig:scalar}
\end{figure}

From the above expressions, one can see that the curvature scalars are influenced by both the string cloud and holonomic correction parameters.

The behavior of these scalar quantities is depict in Figure \ref{fig:scalar} by varying the string cloud parameter $\alpha$, keeping the holonomy correction parameter $\ell$ and the BH mass $M$ fixed.

\section{Null Geodesic and Lensing Phenomena }\label{sec:2}

Null geodesics describe the trajectories followed by massless particles-such as photons as they propagate through a gravitational field generated by curved space-time geometry. These paths are essential in understanding the behavior of light near compact astrophysical objects like BHs. One of the most striking consequences of curved space-time on null geodesics is gravitational lensing (GL), which refers to the bending of light rays as they traverse regions of intense gravitational fields. This phenomenon provided one of the earliest confirmations of Einstein’s general theory of relativity, most notably during the 1919 solar eclipse observations \cite{varios3}. Since then, gravitational lensing has evolved into a powerful observational and theoretical tool in modern cosmology and astrophysics, aiding in the study of dark matter, dark energy, and the large-scale structure of the Universe \cite{AA1,AA2,AA4,AA5}.

Given that the spacetime under consideration (\ref{aa3}) is static and spherically symmetric, the analysis of geodesics can be significantly simplified by restricting the motion of test particles to the equatorial plane, defined by  \( \theta = \pi/2 \). This restriction does not lead to any loss of generality due to the symmetry of the system. To examine particle motion in curved spacetime, we employ the Lagrangian formalism, which allows us to derive the equations of motion from a variational principle. For the given metric, the Lagrangian density function $\mathcal{L}=\frac{1}{2}\,g_{\mu\nu}\,\frac{dx^{\mu}}{d\lambda}\,\frac{dx^{\nu}}{d\lambda}$, where dot represents an ordinary derivative w. r. to an affine parameter along the geodesic and $g_{\mu\nu}$ denotes the components of the metric tensor. This formulation serves as the starting point for deriving the geodesic equations through the Euler-Lagrange equations. References relevant to this analysis include \cite{Kala:2020prt,Kala:2020viz,Kala:2021ppi,Kala:2022uog,Vishvakarma:2024icz,Kala:2024fvg,Pantig:2024lpg,Kala:2025xnb,Roy:2025hdw,Kukreti:2025rzn,Kala:2025fld}, which discuss similar methodologies applied to various modified gravity BH space-times. 

Therefore, for $\theta = \frac{\pi}{2}$, the Lagrangian $\mathcal{L}$ using metric (\ref{aa3}) becomes:
\begin{eqnarray}\label{bb1}
	\mathcal{L}=-\bigg(1-\alpha-\frac{2M}{r}\bigg)\bigg(	\frac{dt}{d\lambda}\bigg)^2+\frac{r}{r-\ell}\bigg(1-\alpha-\frac{2M}{r}\bigg)^{-1}\bigg(\frac{dr}{d\lambda}\bigg)^2+r^2\bigg(\frac{d\phi}{d\lambda}\bigg)^2.
\end{eqnarray}
The corresponding Euler-Lagrange equation for the coordinates $t$ and $\phi$ leads to the following conserved quantities
\begin{equation}\label{bb2}
	\mathrm{E}=\bigg(1-\alpha-\frac{2M}{r}\bigg)\bigg(\frac{dt}{d\lambda}\bigg) \ ,
\end{equation}
and
\begin{equation}\label{bb3}
	\mathrm{L}=r^2\frac{d\phi}{d\lambda} \ .
\end{equation}
which can be understood as energy and angular momentum. Replacing (\ref{bb2}) and (\ref{bb3}) into (\ref{bb1}) and considering null geodesics, where $\mathcal{L}=0 $, we find
\begin{equation}\label{bb4}
	\frac{r}{r-\ell}\bigg(\frac{dr}{d\lambda}\bigg)^2=\mathrm{E}^2-\frac{\mathrm{L}^2}{r^2}\bigg(1-\alpha-\frac {2M}{r}\bigg) \ .
\end{equation}
Eq.(\ref{bb4}) can be seen as describing the dynamics of a classical particle of energy $\mathrm{E}$ subject to an effective potential
\begin{equation}\label{bb5}
	V_\text{eff}=\frac{\mathrm{L}^2}{r^2}\bigg(1-\alpha-\frac{2M}{r}\bigg) \ .
\end{equation}

\begin{figure}[ht!]
\centering
\includegraphics[width=0.4\textwidth]{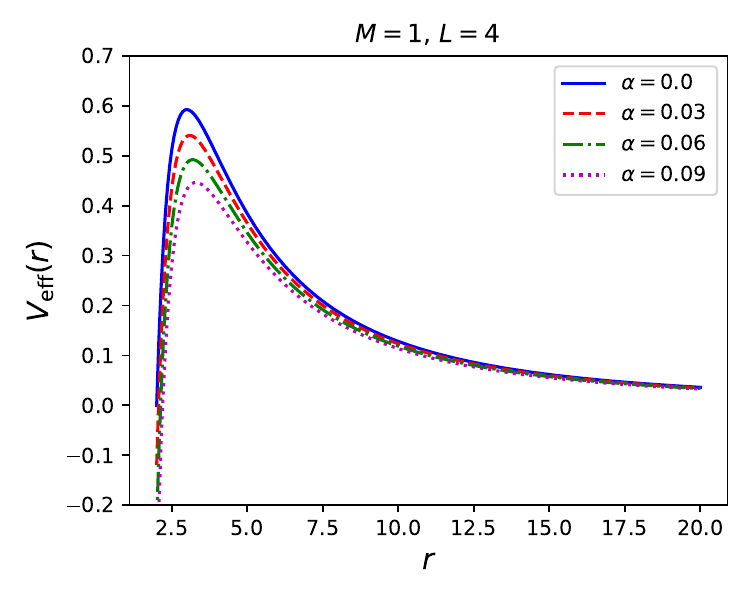}\qquad
\includegraphics[width=0.4\textwidth]{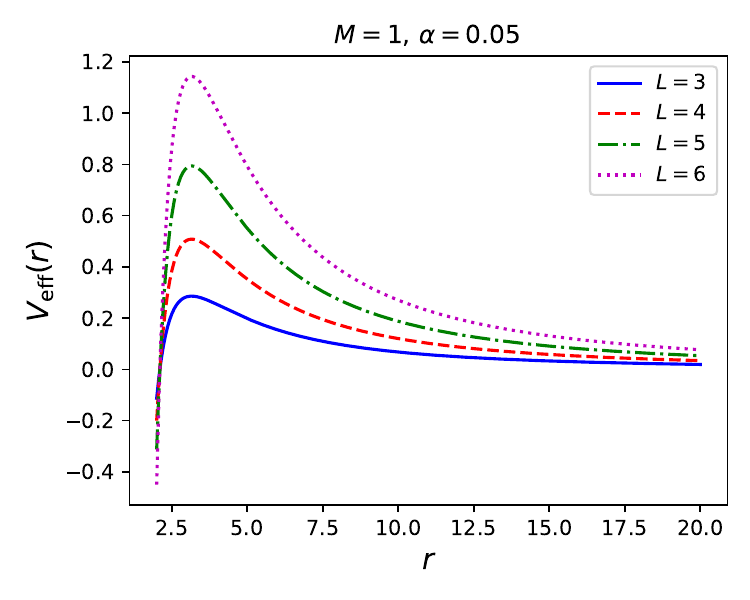}\\
(a) \hspace{6cm} (b)
\caption{Variation of effective potential with radial distance for different values of $\alpha$ and $L$.} \label{VEfffig}
\end{figure} 

The variation of the effective potential with radial distance for different values of the string cloud parameter $\alpha$ and the angular momentum $L$ of massless particles is depicted in Fig.~\ref{VEfffig}. It is clearly evident that increasing either $\alpha$ or $L$ leads to a decrease in the height of the effective potential barrier. Physically, this means that photons require less energy to overcome the potential barrier and move away from the vicinity of the black hole. Furthermore, the effective potential exhibits only a single maximum and no minimum. This feature corresponds to the existence of only one unstable circular photon orbit (the photon sphere). The absence of a potential minimum indicates that no stable circular photon orbits are allowed in this spacetime. Any small perturbation from the maximum causes the photon either to fall into the black hole or escape to infinity. Thus, the effective potential profile encodes the dynamical instability of photon motion around the black hole.  

At the turning point, $r=r_0$, we have $dr/d\lambda=0$. In that case, $V_\text{eff}(r_0)=E^2$, which leads to the following expression
\begin{equation}\label{bb6}
	\frac{1}{\beta^2}=\frac{1}{r_{0}^2}\bigg(1-\alpha-\frac{2M}{r_{0}}\bigg) \ .
\end{equation}
Where $\beta(r_{0})=\frac{\mathrm{L}}{\mathrm{E}}$ is the critical impact parameter. 

\begin{figure} [ht!]
\centering
\includegraphics[width=0.4\textwidth]{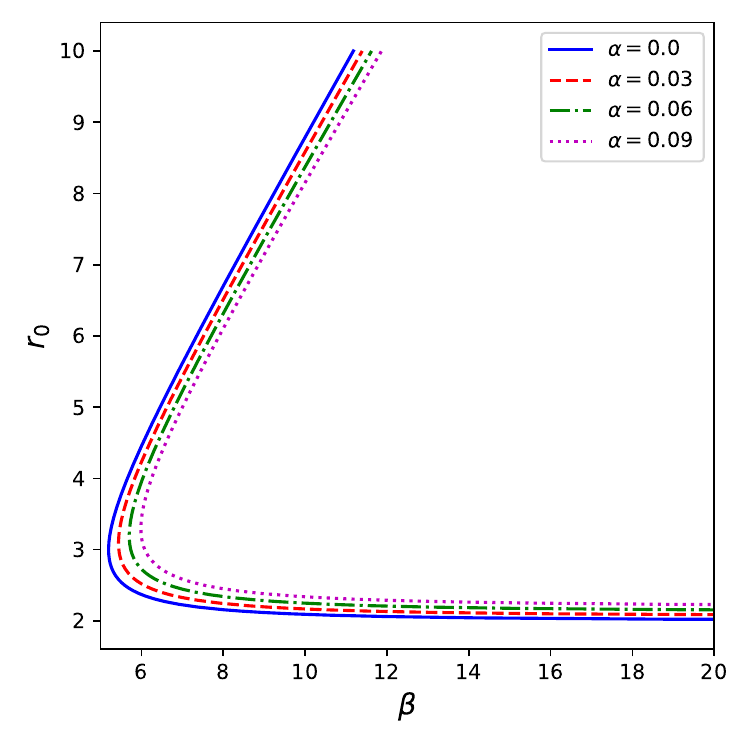} 
\caption{\footnotesize Variation of distance of closest approach as a function of impact parameter for different values of $\alpha$. Here we consider $M=1$.} \label{betar0}
\end{figure}

The distance of closest approach $r_{0}$ corresponds to the minimum radial distance attained by a photon during its trajectory around the black hole before escaping back to infinity. Fig.~\ref{betar0} shows the variation of $r_{0}$ as a function of the impact parameter $\beta$ for different values of the string cloud parameter $\alpha$, with $M=1$. It is observed that $r_{0}$ increases monotonically with $\beta$, and for a fixed $\beta$, larger values of $\alpha$ correspond to smaller values of $r_{0}$. Physically, this indicates that the presence of the string cloud allows photons to penetrate deeper into the gravitational field of the black hole for the same impact parameter, highlighting the role of $\alpha$ in modifying photon trajectories.  

The equation of orbit using Eqs. (\ref{bb2}) and (\ref{bb4}), we find
\begin{equation}\label{bb7}
	\bigg(1-\frac{\ell}{r}\bigg)^{-1}\,\Big(\frac{1}{r^2}\frac{d\phi}{dr}\Big)^2=\frac{1}{\beta^2}-\frac{1}{r^2}\,\bigg(1-\alpha-\frac{2M}{r}\bigg).
\end{equation}
Introducing the following variable change $u=\frac{1}{r}$, from which, we have $dr=-\frac{du}{u^2}$. Therefore, in terms of $u$, (\ref{bb8}) becomes
\begin{equation}\label{bb8}
	\frac{du}{d\phi}=\pm\,\sqrt{(1-\ell\,u)\,\left[\frac{1}{\beta^2}-u^2\,(1-\alpha-2\,M\,u)\right]}. 
\end{equation}

In order to evaluate the integral, we first determine the critical value of the closest approach. Comparing Eq.~\eqref{bb8} with the condition 
\begin{equation}
\left(\frac{du}{d\phi}\right)^{2} - B(u) = 0,\label{bb9}    
\end{equation}

where 
\begin{equation}
    B(u) = (1-\ell\,u)\,\left[\frac{1}{\beta^2}-u^2\,(1-\alpha-2\,M\,u)\right],\label{bb10}
\end{equation}
we note that the turning point of the trajectory (i.e., the closest approach) occurs when 
\(\tfrac{du}{d\phi}=0\). 
Denoting this point as \(u = 1/r_{ps}\), we obtain the critical value of the impact parameter for 
circular photon orbits
\begin{equation}
    \beta_{sc} = \sqrt{\frac{r_{ps}^3}{(1-\alpha)\,r_{ps}-2M}}.\label{bb11}
\end{equation}
According to the circular orbit condition, i.e., by setting $B(u) = 0$ and solving the corresponding equation, we obtain three distinct roots. Among them, one root is real ($u_{1}$), while the other two ($u_{2}$ and $u_{3}$) satisfy the ordering $u_{3} > u_{2} > u_{1}$. These roots are explicitly given by~\cite{Kukreti:2025rzn},
\begin{equation}
    \begin{aligned}
u_{1} &= \frac{\,r_{0}-\alpha r_{0}-2M
 - \sqrt{\bigl(r_{0}-\alpha r_{0}-2M\bigr)\,\bigl(r_{0}-\alpha r_{0}+6M\bigr)}}
 {4M\,r_{0}}, \\[8pt]
u_{2} &= \frac{1}{r_{0}}, \\[8pt]
u_{3} &= \frac{\,r_{0}-\alpha r_{0}-2M 
 + \sqrt{\bigl(r_{0}-\alpha r_{0}-2M\bigr)\,\bigl(r_{0}-\alpha r_{0}+6M\bigr)}}
 {4M\,r_{0}}.\label{bb12}
\end{aligned}
\end{equation}
In addition to these three roots, the presence of the factor $(1 - \ell u)$ in the denominator of Eq.~(\ref{bb10}) introduces another root, which we denote as $u_{+} = \tfrac{1}{\ell}$. This additional modifies the structure of the integrand in the deflection angle calculation. Consequently, the integral can be recast in the form
\begin{equation}
    \Delta \phi = \sqrt{\frac{2}{M \ell}} \left[ \int^{u_2}_{u_1}\frac{du}{\sqrt{(u_{+}-u)(u-u_1)(u-u_2)(u-u_3)}}
    - \int^{0}_{u_1}\frac{du}{\sqrt{(u_{+}-u)(u-u_1)(u-u_2)(u-u_3)}} \right] ,\label{bb13}
\end{equation}
where the turning points $u_{1},u_{2},u_{3}$ satisfy the ordering $u_{3}>u_{2}>u_{1}$, with $u_{+}>u_{3}$.
To evaluate Eq.~\eqref{bb13}, we adopt the method introduced by Iyer and collaborators for reducing quartic integrals to elliptic form \cite{Iyer:2009wa}. Introducing the substitution  
\begin{equation}
    \sin^{2}\psi = \frac{(u-u_{1})(u_{+}-u_{2})}{(u-u_{2})(u_{+}-u_{1})},\label{bb14}
\end{equation}
the integral reduces to the standard elliptic form. Defining  
\begin{equation}
    A = u_{+}-u_{1}, \qquad D = u_{3}-u_{2}, \qquad
    k^{2} = \frac{(u_{+}-u_{2})(u_{3}-u_{1})}{(u_{+}-u_{1})(u_{3}-u_{2})},\label{bb15}
\end{equation}
we obtain  
\begin{equation}
    \int \frac{du}{\sqrt{(u_{+}-u)(u-u_{1})(u-u_{2})(u-u_{3})}}
    = \frac{2}{\sqrt{A D}} \, F(\psi(u),k),\label{bb16}
\end{equation}
where $F(\psi,k)$ is the incomplete elliptic integral of the first kind \cite{Byrd:1971bey},
\begin{equation}
    F(\psi,k) = \int_0^\psi \frac{d\theta}{\sqrt{1-k^2\sin^2\theta}} .\label{bb17}
\end{equation}
Evaluating Eq.~\eqref{bb13} at the integration limits, the exact solution becomes
\begin{equation}
    \Delta \phi
    = \frac{\sqrt{2}}{\sqrt{M\ell \,(u_{+}-u_{1})(u_{3}-u_{2})}}
    \left[ K(k) \;-\; F\!\left(\arcsin \sqrt{\frac{u_{1}(u_{+}-u_{2})}{u_{2}(u_{+}-u_{1})}},\,k\right)\right],
    \label{bb18}
\end{equation}
where $K(k)=F(\tfrac{\pi}{2},k)$ is the complete elliptic integral of the first kind. Now, using Eq.~(\ref{bb18}) the deflection angle is given by
\begin{equation}
    \delta \phi
    = \frac{4}{\sqrt{2M\ell \,(u_{+}-u_{1})(u_{3}-u_{2})}}
    \left[ K(k) \;-\; F\!\left(\arcsin \sqrt{\frac{u_{1}(u_{+}-u_{2})}{u_{2}(u_{+}-u_{1})}},\,k\right)\right].
    \label{bb19}
\end{equation}
Here, $K(k)=F(\tfrac{\pi}{2},k)$ is the complete elliptic integral of the first kind.\\

\begin{figure}[ht!]
\centering
\includegraphics[width=0.4\textwidth]{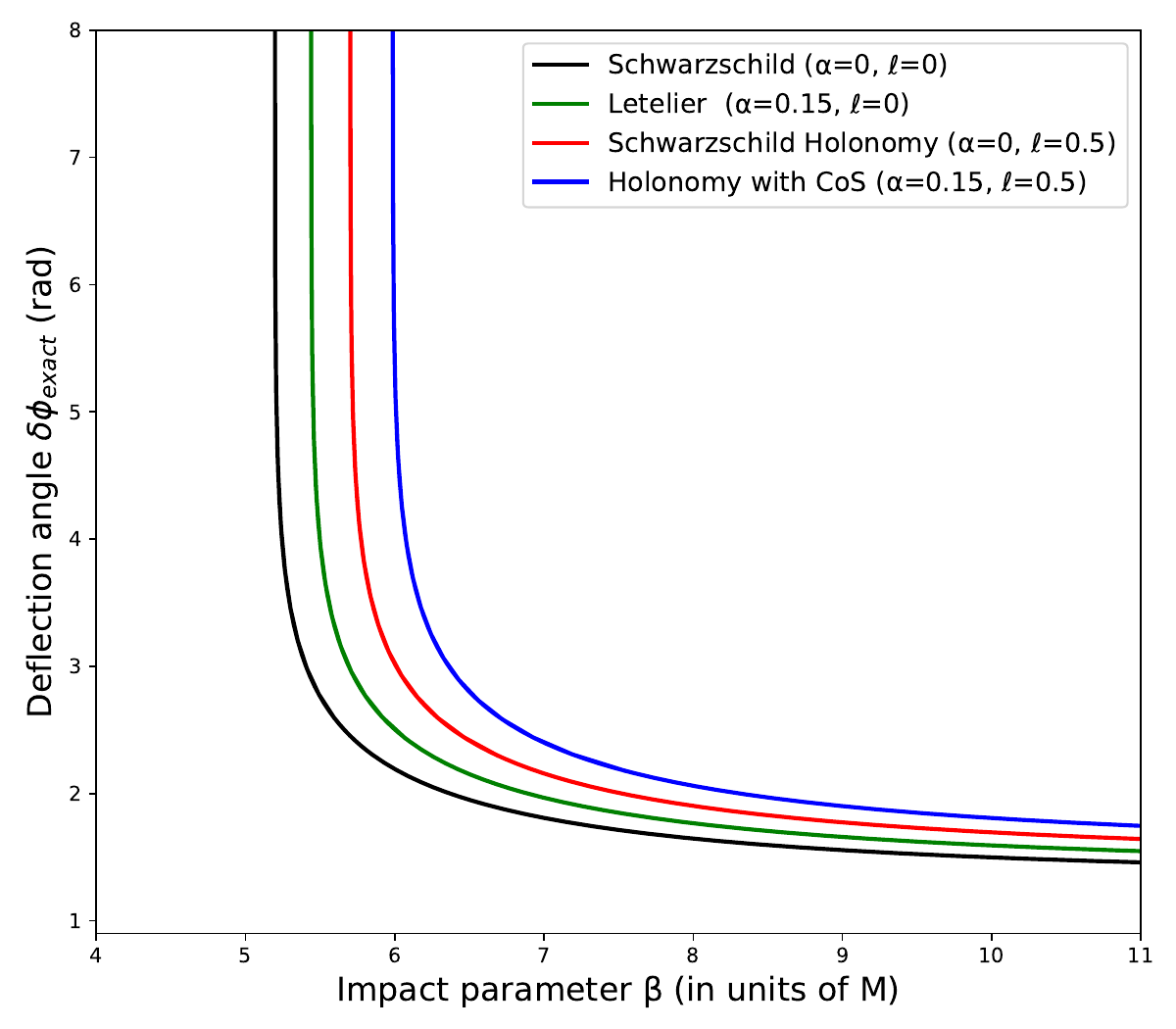}\qquad
\includegraphics[width=0.4\textwidth]{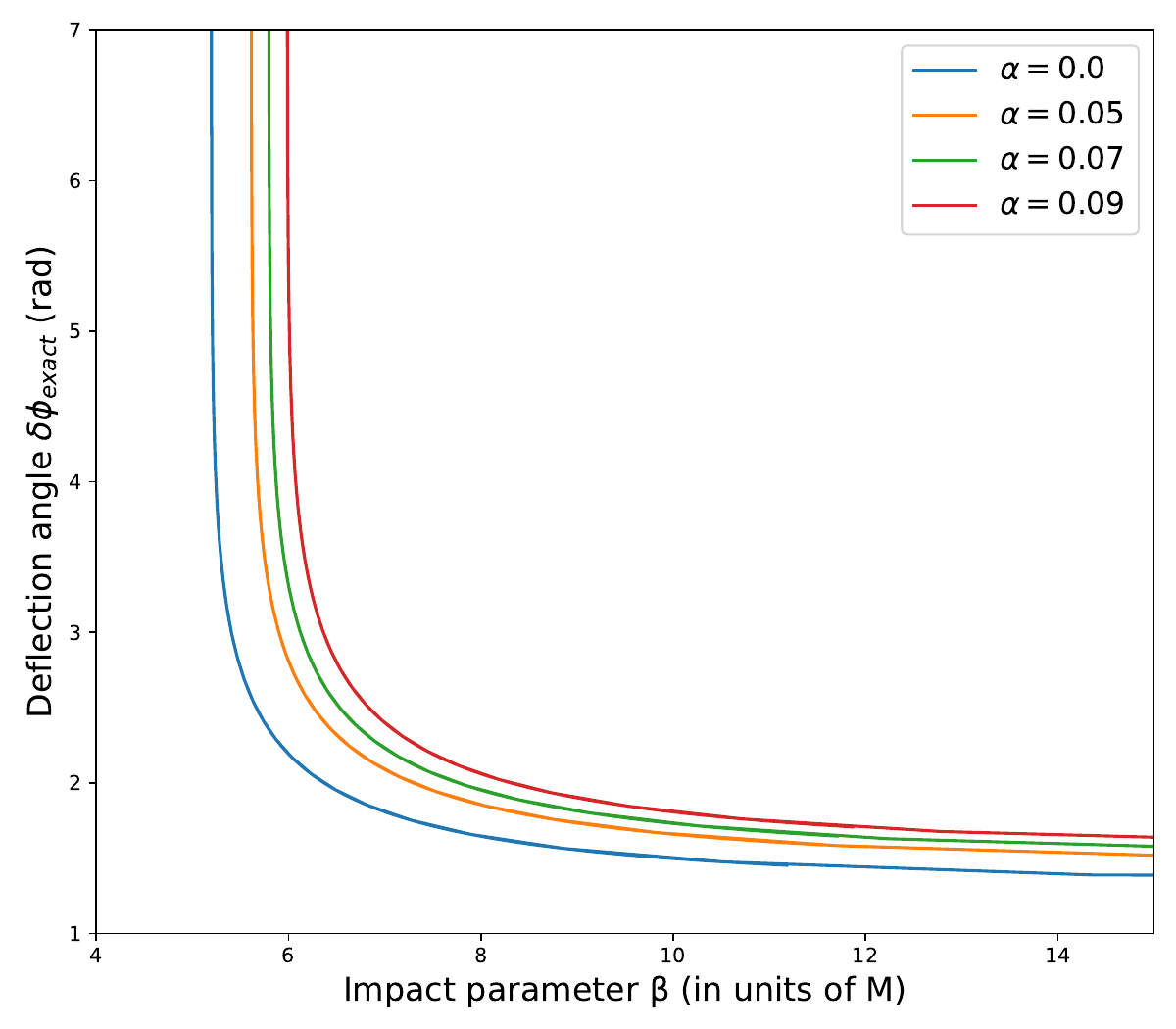}\\
(a) \hspace{6cm} (b)
\caption{ \footnotesize (a) Comparison of the exact deflection angle of the Holonomy-corrected Schwarzschild black hole with other well-known black hole solutions, showing how it reduces to the standard cases in specific limits. (b) Exact deflection angle of the Holonomy-corrected Schwarzschild black hole with absence of the LQG parameter and for different values of the CoS parameter. } \label{fig:defangle01}
\end{figure}

\noindent Using the exact expression for the deflection angle, we compared the Holonomy-corrected Schwarzschild black hole with other well-known black hole solutions, as shown in Fig.~\ref{fig:defangle01}(a), illustrating how it reduces to standard cases in specific limits. We find that the inclusion of both LQG and CoS parameters increases the deflection angle. In the absence of the LQG parameter, we varied the CoS parameter and obtained results, shown in Fig.\ref{fig:defangle01}(b), in excellent agreement with Soares et al.~\cite{HH4}.

In the weak-field approximation, the photon propagates at a large  distance from the black hole such that the gravitational field can be  treated as a small perturbation. In this regime, we may consistently  assume that both the black hole mass parameter $M \ll 1$ and the LQG parameter $\ell \ll 1$. Under these assumptions, the  deflection angle can be expressed as a perturbative expansion in powers of $\ell$. Retaining terms up to second order in $\ell$, Eq.~(\ref{bb19}) provides an analytic approximation for the bending of light as follows~\cite{Pantig:2024lpg},
\begin{equation}
\delta \phi_{weak} \simeq 
\left( \frac{1}{\sqrt{1-\alpha}} - 1 \right)\pi
+ \frac{4M}{\beta (1-\alpha)^{3/2}}
+ \frac{\ell}{\beta \sqrt{1-\alpha}}
+ \frac{3\pi \ell^2\alpha}{16 \beta^{2} \sqrt{1-\alpha}}
+ \frac{\ell M(3\pi - 4)}{4 \beta^{2} (1-\alpha)^{3/2}} \, + \mathcal{O} (M^{2}, \ell^{2}).\label{bb20}
\end{equation}

In the limit $\alpha=0$ corresponding to the absence of string cloud effects, the deflection angle in the weak field limit obtained in Eq. (\ref{bb20}) reduces as,
\begin{equation}
\delta \phi_{weak} \simeq \frac{4M}{\beta}
+ \frac{\ell}{\beta}
+\frac{3 \pi \ell^2}{16 \beta^{2}}
+ \frac{\ell M(3\pi - 4)}{4 \beta^{2}}\, + \mathcal{O} (M^{2}, \ell^{2}).\label{bb20aa}
\end{equation}
The result obtained above closely resembles those found in holonomy-corrected Schwarzschild BH, as discussed in \cite{HH4}.

Moreover, in the limit $\ell \to 0$, corresponding to the absence of holonomic corrections, the deflection angle obtained in Eq. (\ref{bb20}) reduces as,
\begin{equation}
\delta \phi_{weak} \simeq 
\left( \frac{1}{\sqrt{1-\alpha}} - 1 \right)\pi
+ \frac{4M}{\beta (1-\alpha)^{3/2}}.\label{bb20b}
\end{equation}

Thus, the first term in Eq. (\ref{bb20}) originates from the presence of strings cloud surrounding the BH, representing the influence of this external matter distribution on the geometry of space-time. The second term accounts for the modification of the deflection angle compared to the standard Schwarzschild BH, highlighting the deviation caused by the strings cloud. The remaining terms incorporate the effects of holonomic (quantum gravitational) corrections and strings cloud, indicating how quantum geometry and matter content collectively alter the geodesic structure and lensing properties of the space-time.

The expression (\ref{bb20}) not only facilitates direct comparison with the classical Schwarzschild result in the limit $\ell \to 0$, but also highlights the contribution of higher-order corrections arising from the parameter $\ell$. Such corrections are particularly relevant in assessing the impact of deviations from standard general relativity in the weak-field regime, where observational data (e.g., from gravitational lensing surveys) are most readily available.

\begin{figure}[ht!]
\centering
\includegraphics[width=0.4\textwidth]{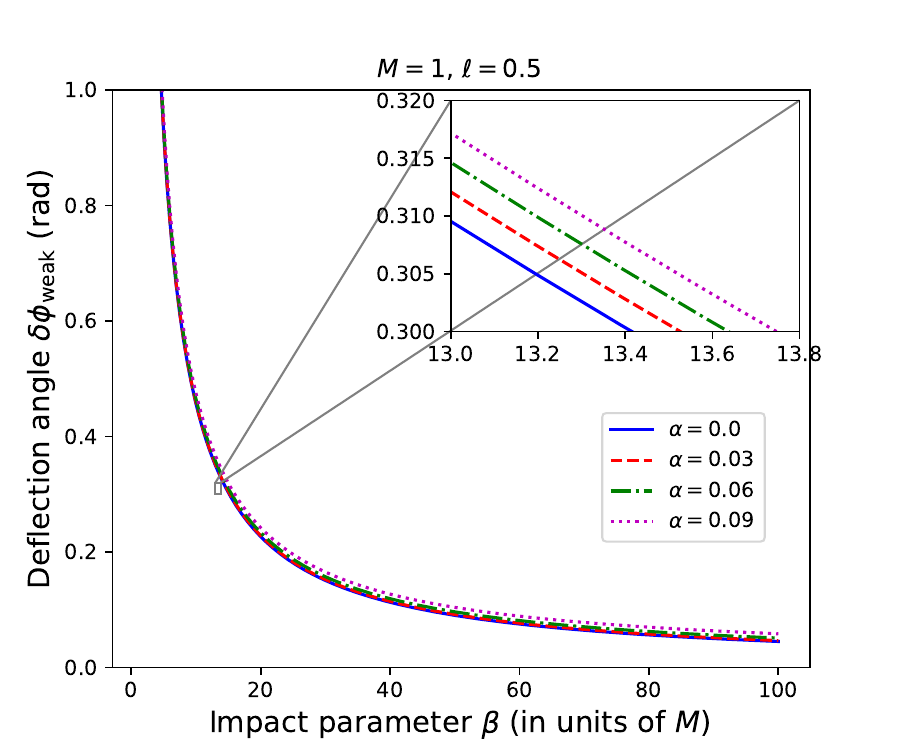}\qquad
\includegraphics[width=0.4\textwidth]{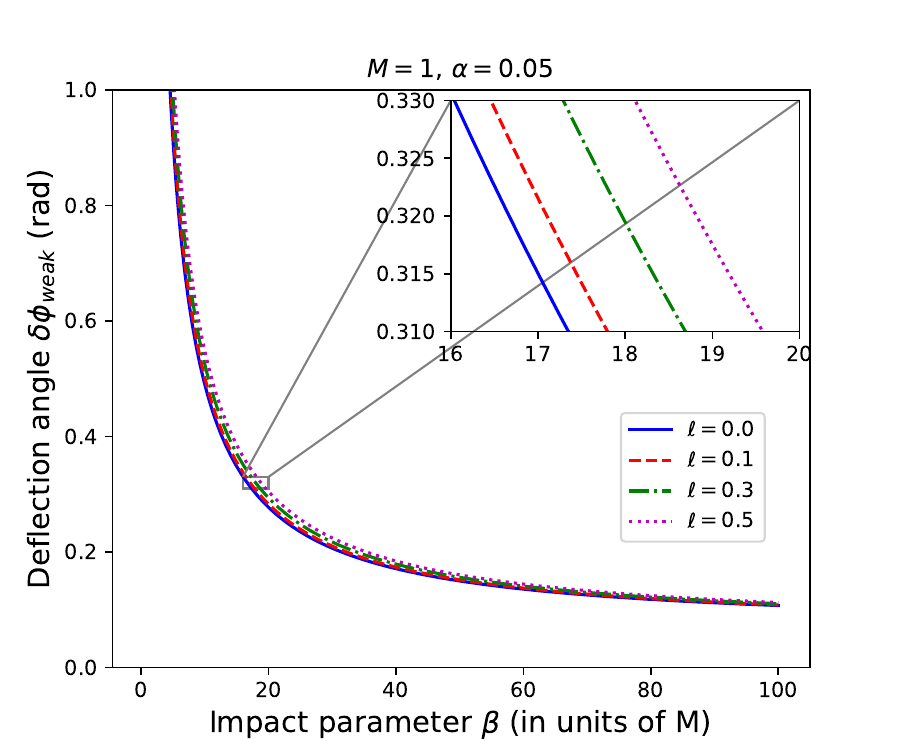}\\
(a) \hspace{6cm} (b)
\caption{Variation of deflection angle as a function of impact parameter for different values of $\alpha$ and $\ell$.} \label{fig:defangle1}
\end{figure}

\begin{figure}[ht!]
\centering
\includegraphics[width=0.6\textwidth]{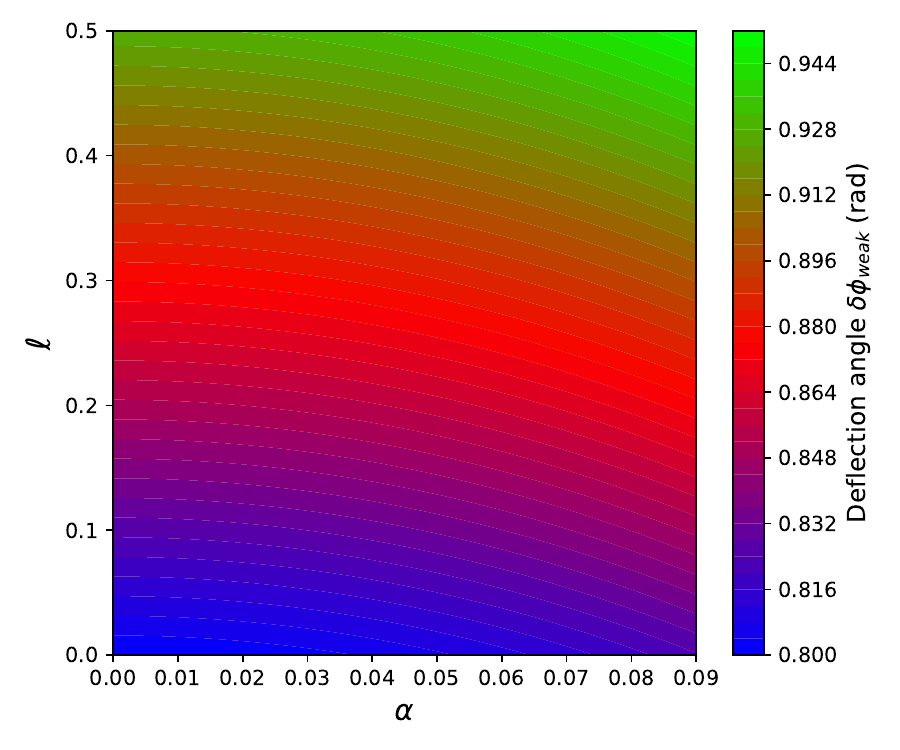}
\caption{Density plot showing the variation of deflection angle as a function of $\alpha$ and $\ell$.} \label{fig:defangle2}
\end{figure}

The variation of the deflection angle as a function of the impact parameter is shown for different values of $\alpha$ (CoS parameter) and $\ell$ (LQG parameter) in Fig.~\ref{fig:defangle1}. As expected, the deflection angle decreases with increasing impact parameter, reflecting the weakening of gravitational influence at larger distances from the black hole. Additionally, the deflection angle increases with higher values of both $\alpha$ and $\ell$, indicating that the modifications introduced by these parameters strengthen the spacetime curvature, leading to stronger light bending. This behavior is further illustrated in the density plot shown in Fig.~\ref{fig:defangle2}, which clearly highlights the regions of enhanced deflection. Physically, the increase in deflection with $\alpha$ and $\ell$ suggests that the CoS and LQG corrections effectively enhance the gravitational lensing signature, which could potentially serve as observational probes to constrain these parameters in strong gravity regimes.

 \section{Lens Equation and Magnification}\label{sec:3}
 
In the weak deflection regime of gravitational lensing, the geometry of the lens system can be described by a simple relation between the angular positions of the source, the image, and the deflection angle. This relation, known as the lens equation, provides the basis for calculating observable quantities such as the Einstein ring radius and the magnification of lensed images. It allows us to connect the theoretical deflection angle obtained from the BH spacetime with measurable astrophysical effects. In this direction, Bozza has proposed an improved tool for approximate gravitational lens equations in a static and spherically BHs~\cite{ref51a}. 

Now, we explore the brightness of the images produced by a holonomy-corrected Schwarzschild BH surrounded by a cloud of strings. Using the lens equation, the combination of angular quantities around the BH can be written as \cite{Schneider:1992bmb}, 
\begin{equation}
\theta D_{s} = \zeta D_{s} + \delta_{\phi} D_{ds}, 
\label{eq:lens}
\end{equation}
where $D_{s}$, $D_{d}$, and $D_{ds}$ denote the distances from the source to the observer, from the lens to the observer, and from the source to the lens, respectively. Here, $\theta$ and $\zeta$ represent the angular position of the image and the source.  

From Eq.~\eqref{eq:lens}, we can rewrite the source angle as  
\begin{equation}
\zeta = \theta - \frac{D_{ds}}{D_{s}} \, \xi(\theta) \, \frac{1}{D_{d}\theta}, 
\label{eq:zeta}
\end{equation}
where $\xi(\theta) = |\delta_{\phi}\,\beta|\,\beta$ and $\beta = D_{d}\theta$. If the image forms a perfect ring, it is identified as the Einstein ring, whose radius is $R_{s} = D_{d}\theta_{E}$. The corresponding Einstein angle due to the spacetime geometry is~\cite{Bozza:2010xqn,Vishvakarma:2024icz}, 
\begin{equation}
\theta_{E} = \sqrt{\frac{2R_{s}D_{ds}}{D_{d}D_{s}}}.
\label{eq:einstein}
\end{equation}

The magnification of the image brightness is then given by~\cite{Al-Badawi:2024dzc},
\begin{equation}
\mu = \frac{I_{\text{tot}}}{I^{\ast}}
= \sum_{k=1}^{j} \left| \frac{\theta_{k}}{\zeta} \frac{d\theta_{k}}{d\zeta} \right|, \quad k=1,2,3.....j,
\label{eq:mag}
\end{equation}
where $I^{\ast}$ and $I_{\text{tot}}$ denote the unlensed brightness of the source and the total brightness of all lensed images, respectively.  

\begin{figure}[ht!]
\centering
\includegraphics[width=0.4\textwidth]{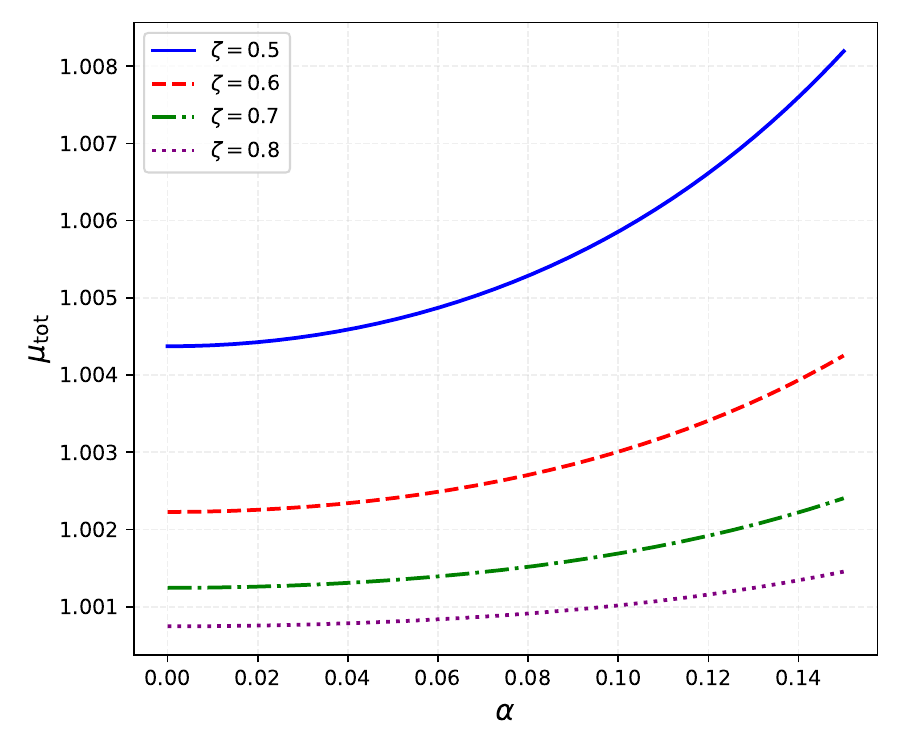}\qquad
\includegraphics[width=0.4\textwidth]{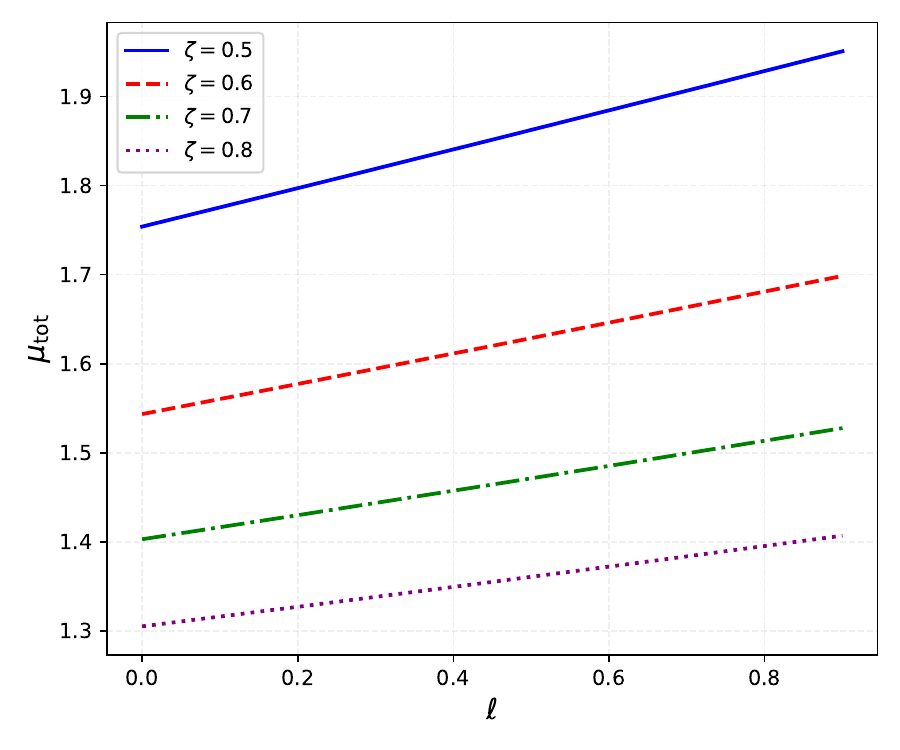}\\
(a) \hspace{6cm} (b)
\caption{Variation of total magnification as a function of $\alpha$ and $\ell$ for different values of $\zeta$. Here we consider, $D_{ds}=10$, $D_{s}=5$ and $\beta=6$.} \label{fig:MAG1}
\end{figure}

\begin{figure}[ht!]
\centering
\includegraphics[width=0.5\textwidth]{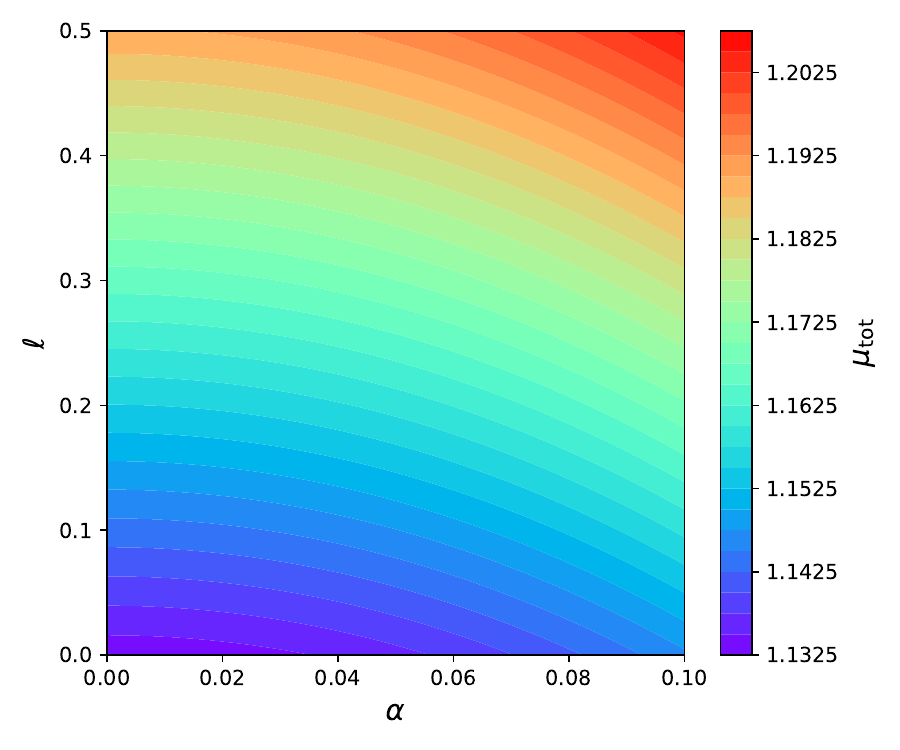}
\caption{Density plot showing the variation of total magnification as a function of $\alpha$ and $\ell$. Here we consider, $D_{ds}=10$, $D_{s}=5$, $\zeta=0.5$ and $\beta=6$.} \label{fig:MAG3}
\end{figure}

The explicit expressions for the magnifications of the two images are   
\begin{equation}
\mu_{+} = \frac{1}{4} \left( \frac{\chi}{\sqrt{\chi^{2} + 4}}  + \frac{\sqrt{\chi^{2} + 4}}{\chi}+2 \right),
\label{eq:muplus}
\end{equation}
\begin{equation}
\mu_{-} =\frac{1}{4} \left( \frac{\chi}{\sqrt{\chi^{2} + 4}}  + \frac{\sqrt{\chi^{2} + 4}}{\chi}-2 \right),
\label{eq:muminus}
\end{equation}
where $\chi = \zeta/\theta_{0}$ is a dimensionless parameter. The total magnification can then be obtained as  
\begin{equation}
\mu_{\text{tot}} = \mu_{+} + \mu_{-} = \frac{\chi^{2} + 2}{\chi \sqrt{\chi^{2} + 4}}.
\label{eq:mutot}
\end{equation}

The variation of the total magnification as a function of $\alpha$ and $\ell$ for different values of $\zeta$ (angular position of the source) is depicted in Fig.~\ref{fig:MAG1}. The total brightness of the lensed images increases with increasing $\alpha$ and $\ell$, indicating that the CoS and LQG parameters enhance the spacetime curvature, thereby strengthening the lensing effect. Notably, for smaller values of the source angular position $\zeta$, the magnification increases sharply, reflecting the well-known fact that light rays passing closer to the line of sight experience stronger gravitational focusing. As $\zeta$ increases, the sharpness of this enhancement gradually decreases, consistent with the weakening of lensing for sources farther from the optical axis. Furthermore, the density plot of the total magnification as a function of $\alpha$ and $\ell$, also shown in Fig.~\ref{fig:MAG3}, clearly illustrates regions of enhanced lensing. Physically, this suggests that observations of magnified images could serve as sensitive probes for detecting or constraining the values of $\alpha$ and $\ell$ in strong-gravity regimes.

\section{Topological Properties of Photon Rings}\label{sec:4}

In this section, we aim to investigate topological properties of photon sphere in a holonomy corrected Letelier BH, focusing particularly on the role of photon spheres (PSs). Photon spheres are regions where null geodesics, or photon orbits, can exist, providing important insights into the stability and structure of BHs. In this work, we following the methodology \cite{ref33,ref34,ref36} and discuss the topological feature of the photon sphere showing the effects of holonomy corrections and string clouds.

The geodesic paths for radial coordinate $r$ using Eq.~(\ref{bb4}) can be re-written as
\begin{equation}
    \left(\frac{dr}{d\lambda}\right)^2+U_\text{eff}=0,\label{cc1}
\end{equation}
where the effective potential as,
\begin{equation}
    U_\text{eff}=\left(1-\frac{\ell}{r}\right)\,\left[\mathrm{E}^2-\frac{\mathrm{L}^2}{r^2}\,f(r)\right].\label{cc2}
\end{equation}

For circular orbits of radius $r=r_0$, the conditions $\frac{dr}{d\lambda}=0$ and $\frac{d^2r}{d\lambda^2}=0$ must be satisfied. The first condition results Eq. (\ref{bb7}), while the second condition gives us the radius of photon sphere satisfying the following relation:
\begin{eqnarray}
    \frac{d}{dr}\left(\frac{f(r)}{r^2}\right)=0.\label{cc3}
\end{eqnarray}
The simplification of the above condition gives the photon sphere radius as,
\begin{equation}
    r_\text{ph}=\frac{3\,M}{1-\alpha}.\label{cc4}
\end{equation}

\begin{table}[ht!]
\centering
\begin{tabular}{|c|c|c|c|c|c|c|c|c|c|}
\hline
$\alpha$ & 0 & 0.05 & 0.10 & 0.15 & 0.20 & 0.25 & 0.30 & 0.35 & 0.40 \\
\hline
$r_{\text{ph}}$ & $3M$ & $3.158M$ & $3.333M$ & $3.529M$ & $3.750M$ & $4.000M$ & $4.286M$ & $4.615M$ & $5M$ \\
\hline
\end{tabular}
\caption{Photon sphere radius $r_{\text{ph}}$ as a function of $\alpha$}
\label{tab:1}
\end{table}

In 1984, Duan carried out a seminal analysis of the intrinsic structure of conserved topological currents in the context of \( SU(2) \) gauge theory. In this study, he introduced the notion of topological flow, associated with point-like systems resembling particles. This foundational work established the groundwork for later developments involving various types of topological currents~\cite{YSD}. As a first step in the formulation, we introduce a general vector field \(\mathbf{v}\), which can be decomposed into two components: the radial component \(v_r\) and the angular component \(v_{\theta}\).

\begin{figure}[ht!]
    \centering
    \includegraphics[width=0.4\linewidth]{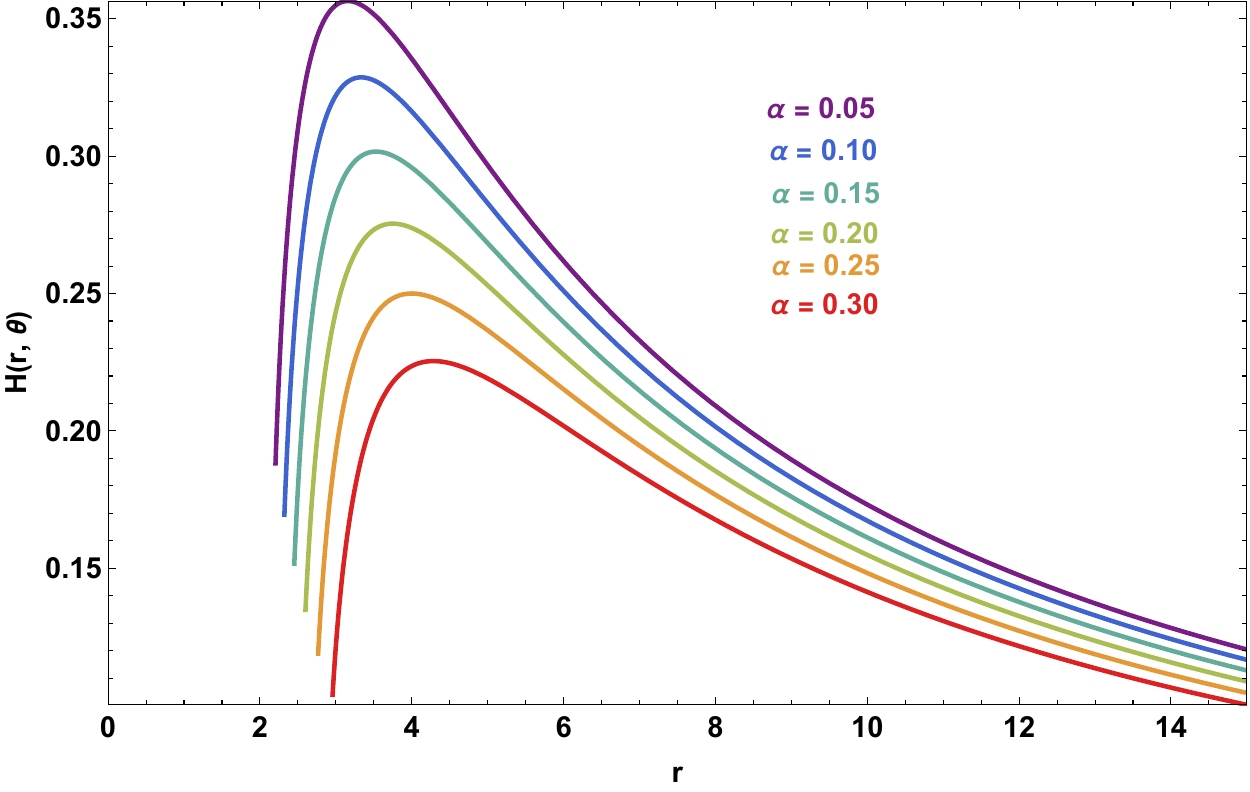}\quad
    \includegraphics[width=0.4\linewidth]{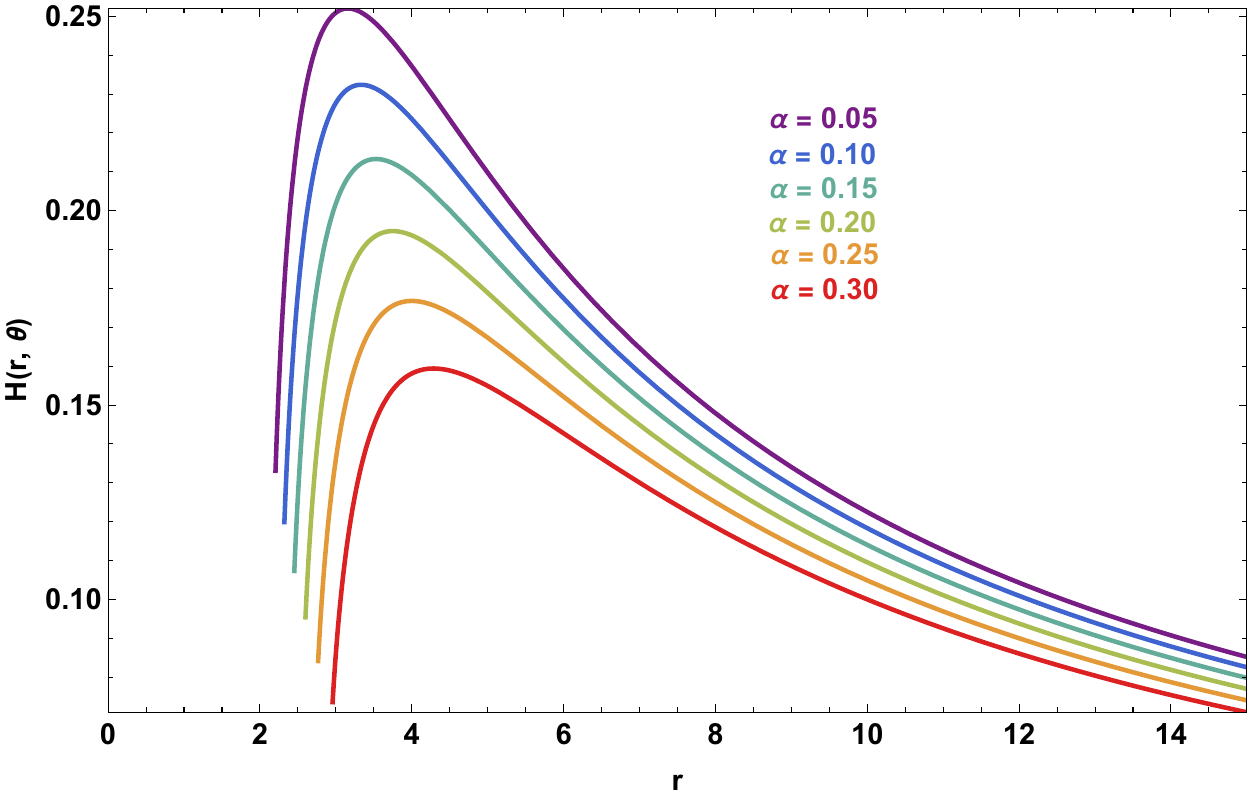}\\
    (a) $\theta=\pi/6$ \hspace{6cm} (b) $\theta=\pi/4$\\ 
    \includegraphics[width=0.4\linewidth]{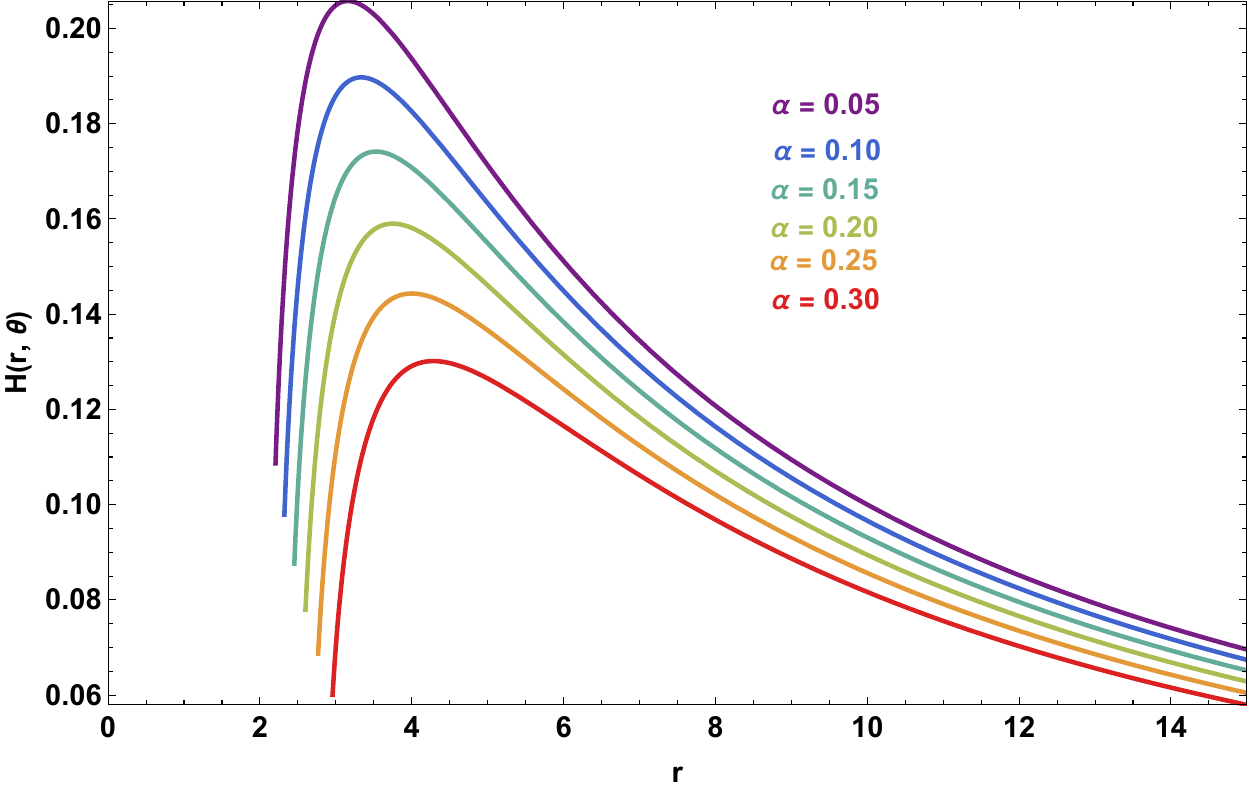}\quad
    \includegraphics[width=0.4\linewidth]{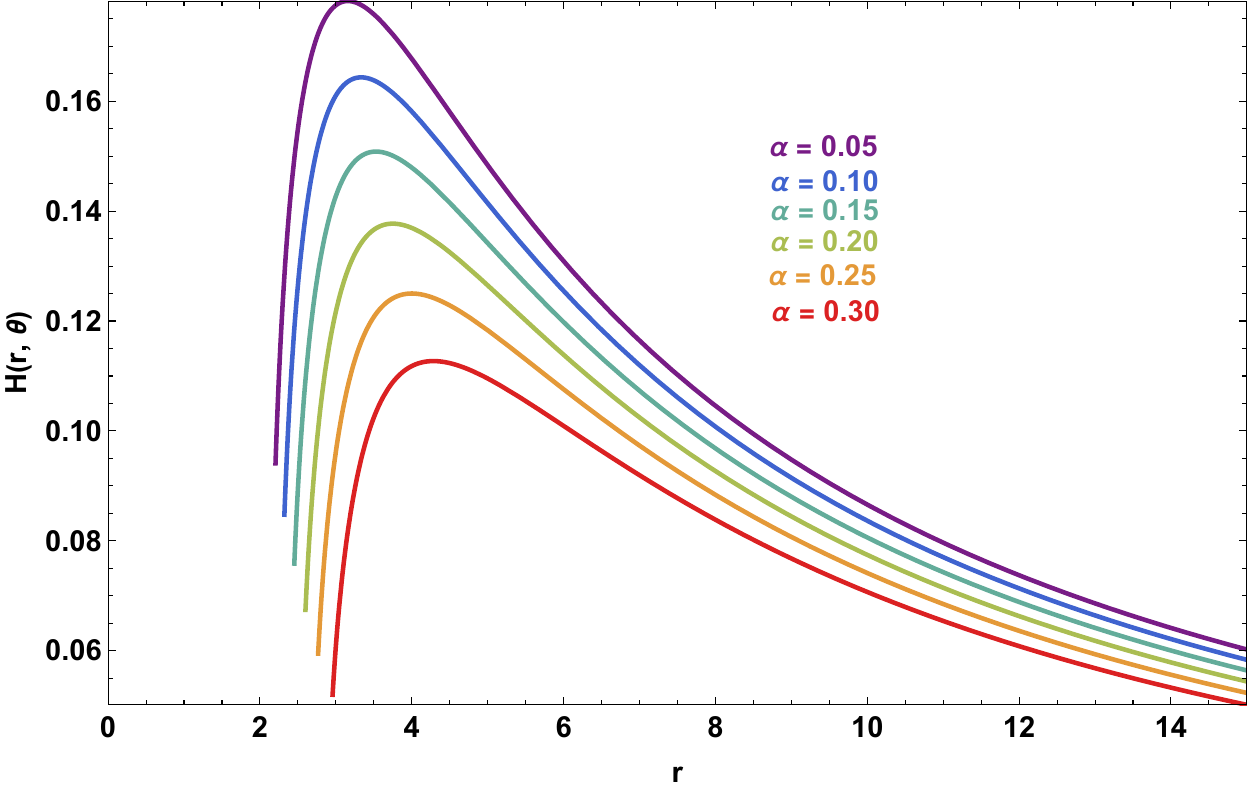}\\
    (c) $\theta=\pi/3$ \hspace{6cm} (d) $\theta=\pi/2$
    \caption{\footnotesize The behavior of the potential function $H(r,\theta)$ by varying CoS parameter $\alpha$ for different $\theta$.}
    \label{fig:vector-potential}
\end{figure}

To investigate the topological properties of light rings, we begin by considering a regular potential expressed through a function as follows \cite{ref33,ref34,ref36}:
\begin{equation}
    H(r,\theta)=\sqrt{-\frac{g_{tt}}{g_{\theta\theta}}}=\frac{\sqrt{1-\alpha-\frac{2M}{r}}}{r\,\sin \theta},\label{cc5}
\end{equation}
where the function $H(r, \theta)$ is regular for $r >\frac{2M}{1-\alpha}$. In Figure \ref{fig:vector-potential}, we depict the potential function $H(r,\theta)$ by varying CoS parameter $\alpha$ for different $\theta=\pi/6,\,\pi/4$ and $\pi/3$, while keeping the mass $M=1$ to be fixed.

The key general vector field ${\bf v}=(v_r\,,\,v_{\theta})$ with the components defined as,
\begin{align}
    v_r&=-\frac{1}{r^2\,\sin\theta}\sqrt{1-\frac{\ell}{r}}\,\left(1-\alpha-\frac{3\,M}{r}\right),\label{cc6}\\
    v_{\theta}&=-\frac{\sqrt{1-\alpha-\frac{2M}{r}}}{r^2}\,\frac{\cot \theta}{\sin \theta}.\label{cc7}
\end{align}
The vector field can be consider as ${\bf v}=v_r+i\,v_{\theta}$. Therefore, the unit vector field (or the normalized vector field) is given by
\begin{equation}
    {\bf n}=(n_r\,,\,n_{\theta})=\frac{{\bf v}}{|{\bf v}|},\label{cc8}
\end{equation}
where $|{\bf v}|=\sqrt{v^2_r+v^2_{\theta}}$ and 
\begin{align}
    n_r&=-\frac{\sqrt{1-\frac{\ell}{r}}\,\left(1-\alpha-\frac{3\,M}{r}\right)}{\sqrt{\left(1-\frac{\ell}{r}\right)\,(\left(1-\alpha-\frac{3\,M}{r}\right)^2+\left(1-\alpha-\frac{2\,M}{r}\right)\,\cot^2 \theta}},\label{cc9}\\
    n_{\theta}&=-\frac{\sqrt{1-\alpha-\frac{2\,M}{r}}\,\cot \theta}{\sqrt{\left(1-\frac{\ell}{r}\right)\,(\left(1-\alpha-\frac{3\,M}{r}\right)^2+\left(1-\alpha-\frac{2\,M}{r}\right)\,\cot^2 \theta}}.\label{cc10}
\end{align}

\begin{figure}[ht!]
    \centering
    \includegraphics[width=0.4\linewidth]{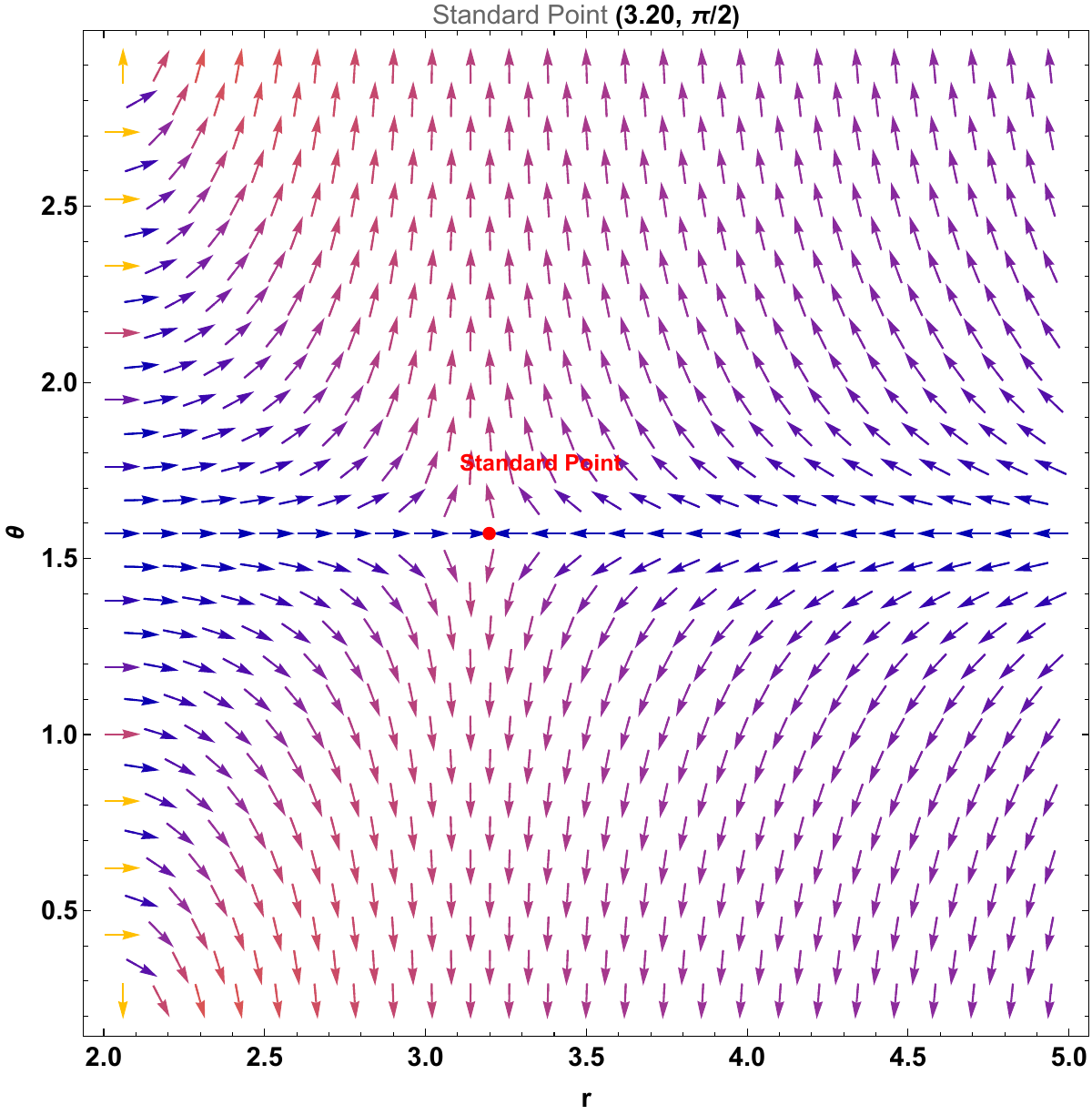}\qquad
    \includegraphics[width=0.4\linewidth]{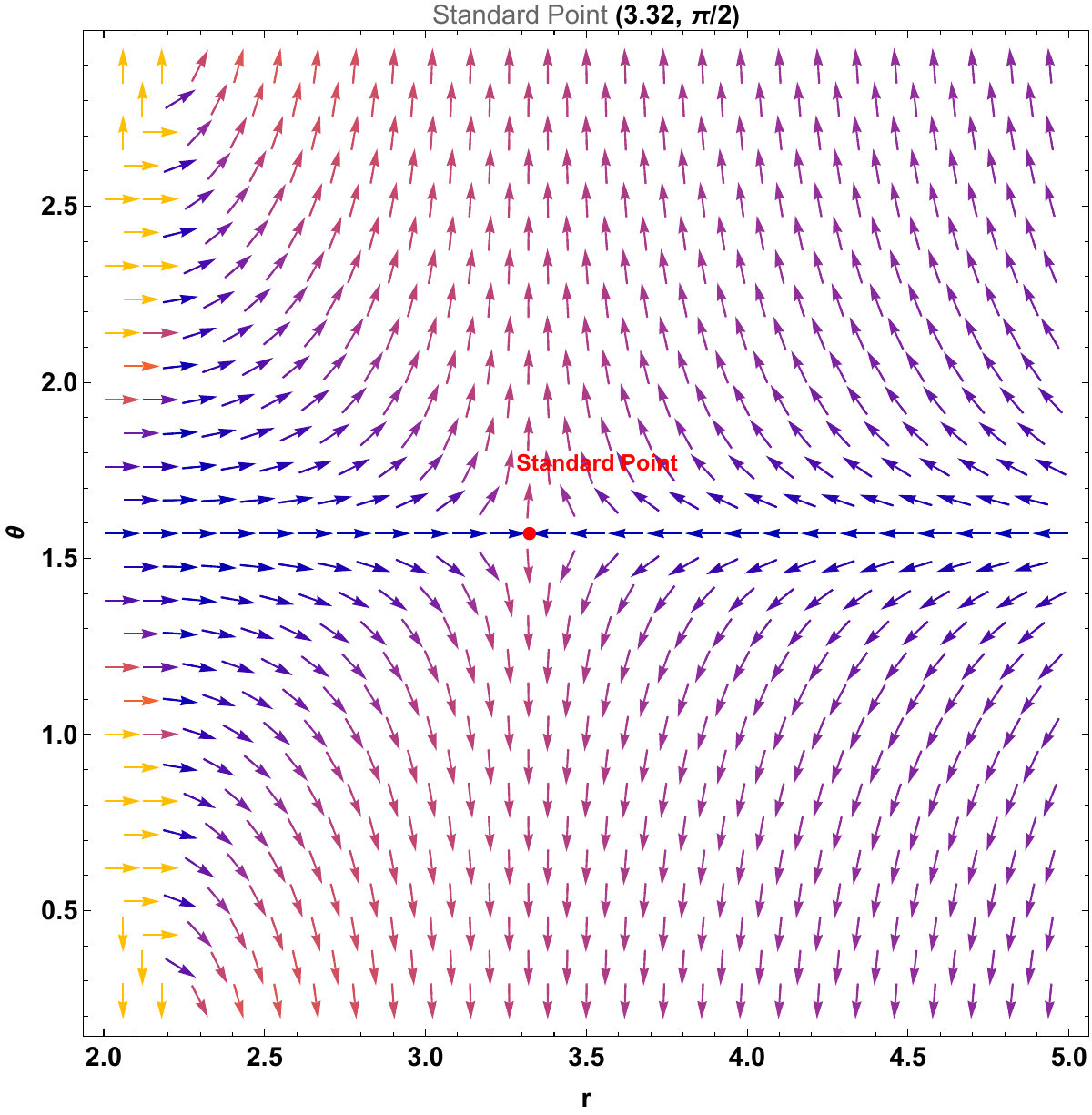}\\
    (a) $\alpha=0.05$ \hspace{6cm} (b) $\alpha=0.10$\\
    \includegraphics[width=0.4\linewidth]{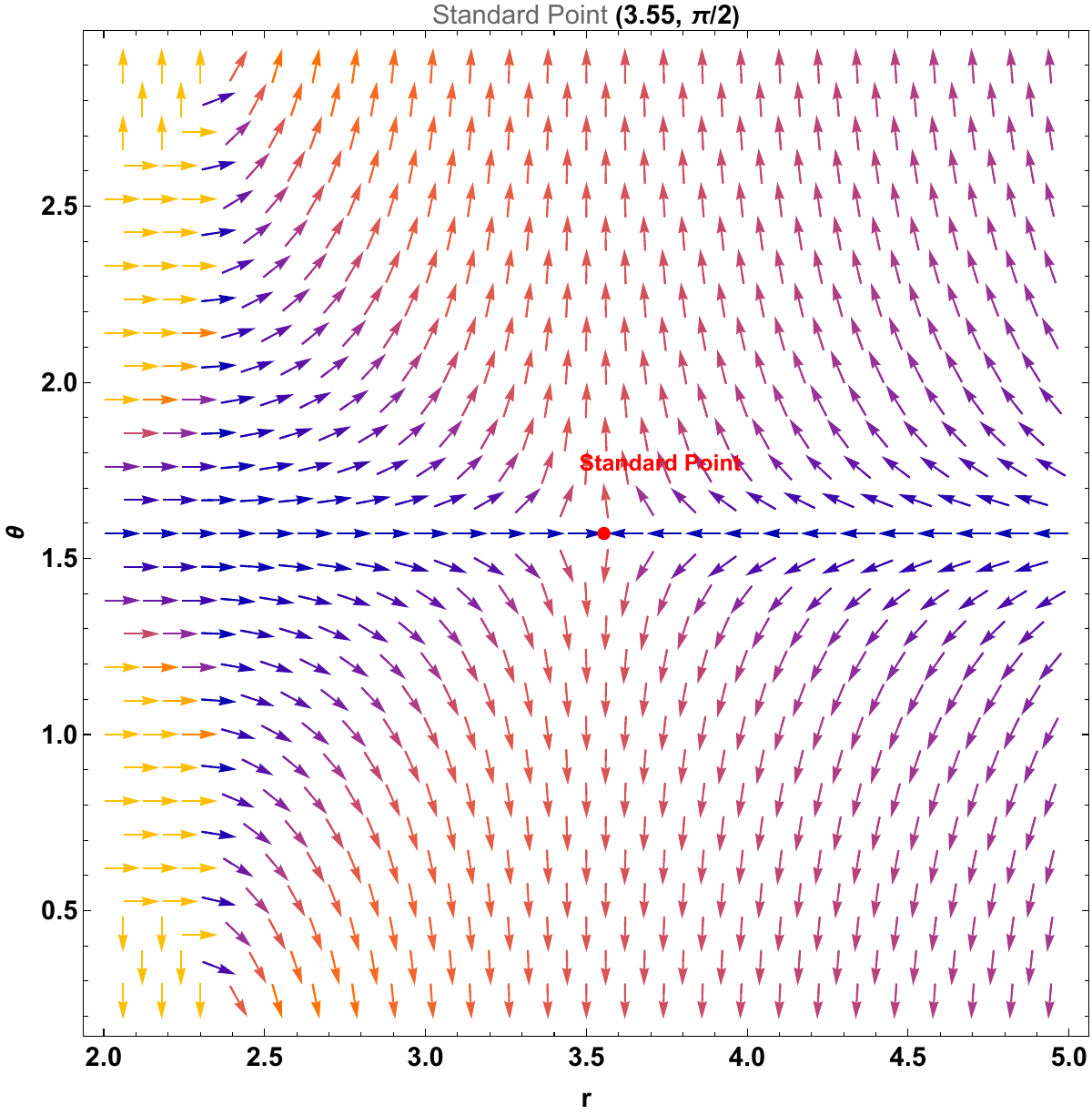}\qquad
    \includegraphics[width=0.4\linewidth]{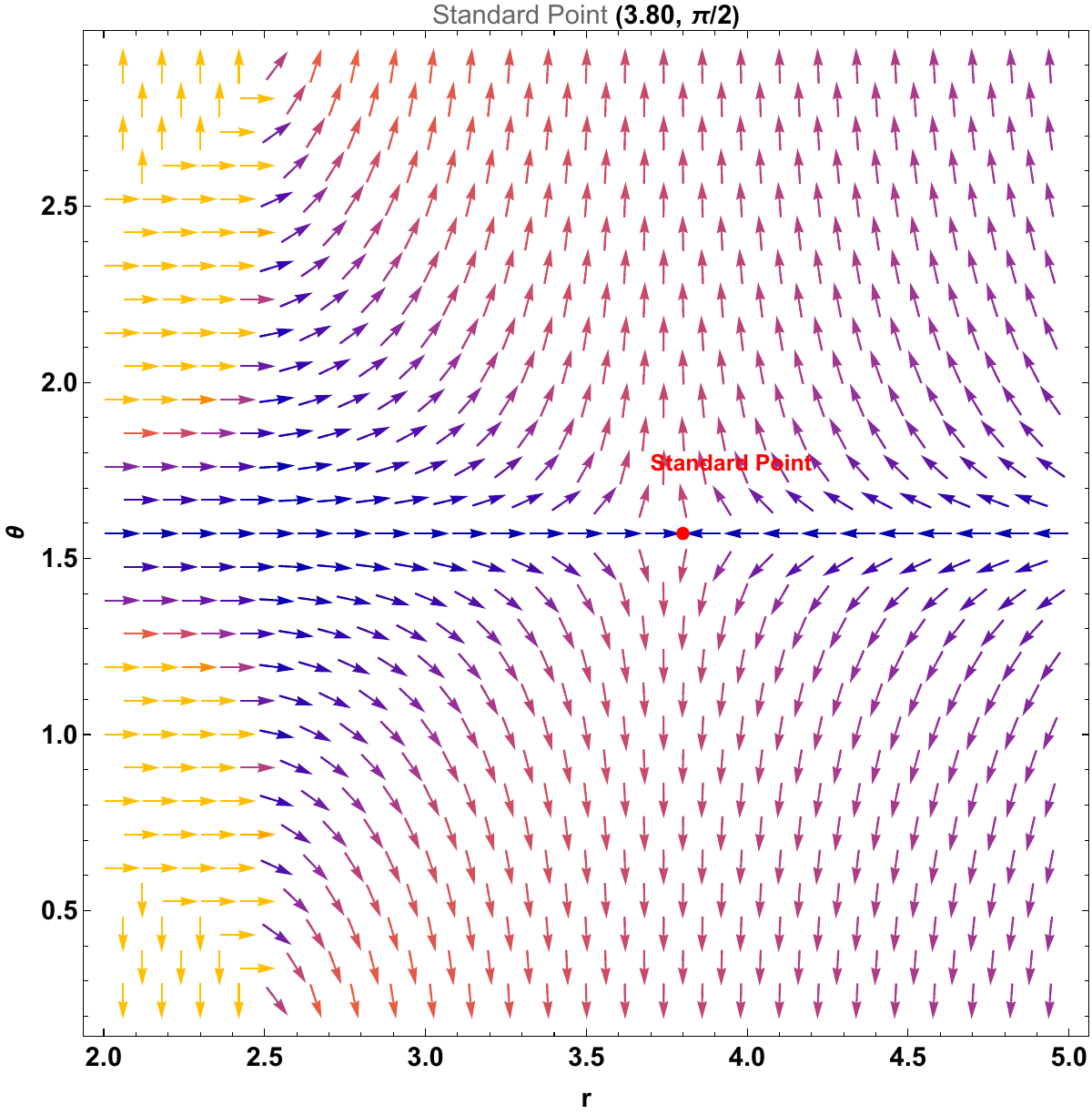}\\
     (c) $\alpha=0.15$ \hspace{6cm} (d) $\alpha=0.20$
    \caption{\footnotesize The arrows represent the normalized vector ${\bf n}$ on a portion of the $r-\theta$ plane for the BH solution with $M=1$ and $\ell=0.1$. The red dot is at $(r,\theta)=(r_0,\pi/2)$ with $r_0=r_\text{ph}\pm b$ where $b$ is the correction term. All panels corresponding to the holonomy-corrected Letelier BH solution.}
    \label{fig:vector}
\end{figure}

From the above analysis, it becomes evident that the normalized vector field $\mathbf{n}$ is significantly influenced by three key parameters: the cloud of strings parameter $\alpha$, the holonomy correction parameter $\ell$, and the BH mass $M$. These dependencies become particularly relevant when evaluated at specific angular coordinates $\theta \neq 0$, especially near the equatorial plane. 

In Figure~\ref{fig:vector}, we visualize the behavior of the normalized vector field $\mathbf{n}$ over a segment of the $r{-}\theta$ plane. The plot corresponds to a BH configuration with fixed mass $M = 1$ and holonomy correction parameter $\ell = 0.1$. A red dot is marked at the location $(r, \theta) = (r_0, \pi/2)$, where the radial coordinate $r_0$ is taken as $r_{\text{ph}} \pm b$. Here, $r_{\text{ph}}$ denotes the photon sphere radius in the classical limit, and $b$ is a small correction term arising from the holonomy-induced modifications to the BH geometry.

It is important to emphasize that, from a mathematical standpoint, the vector field $\mathbf{v}$ vanishes exactly at the point $(r, \theta) = (r_{\text{ph}}, \pi/2)$, as this corresponds to a critical point of the field. However, due to the presence of quantum gravity corrections (introduced via the holonomy parameter $\ell$), the vanishing behavior of $\mathbf{n}$ is slightly deformed in its graphical representation. This results in a visually non-zero, yet theoretically negligible, vector magnitude at that location. Such deviations in the vector field structure highlight the non-trivial effects of holonomy corrections on the topological properties of the BH spacetime. These modifications could have further implications on the stability of photon orbits, the nature of light rings, and potentially on observable signatures in BH shadow imaging. 

\begin{figure}[ht!]
    \centering
    \includegraphics[width=0.4\linewidth]{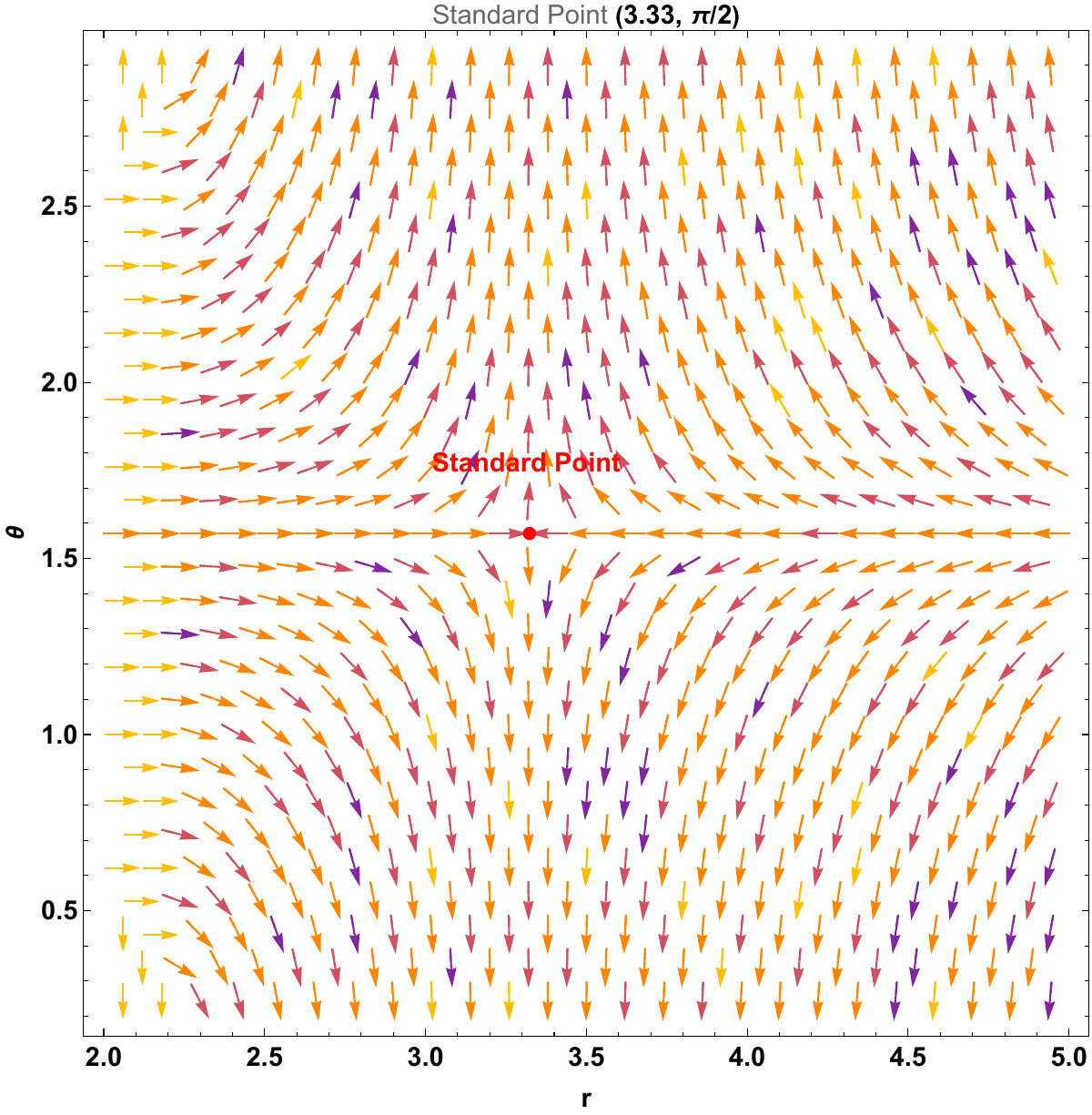}\qquad
    \includegraphics[width=0.4\linewidth]{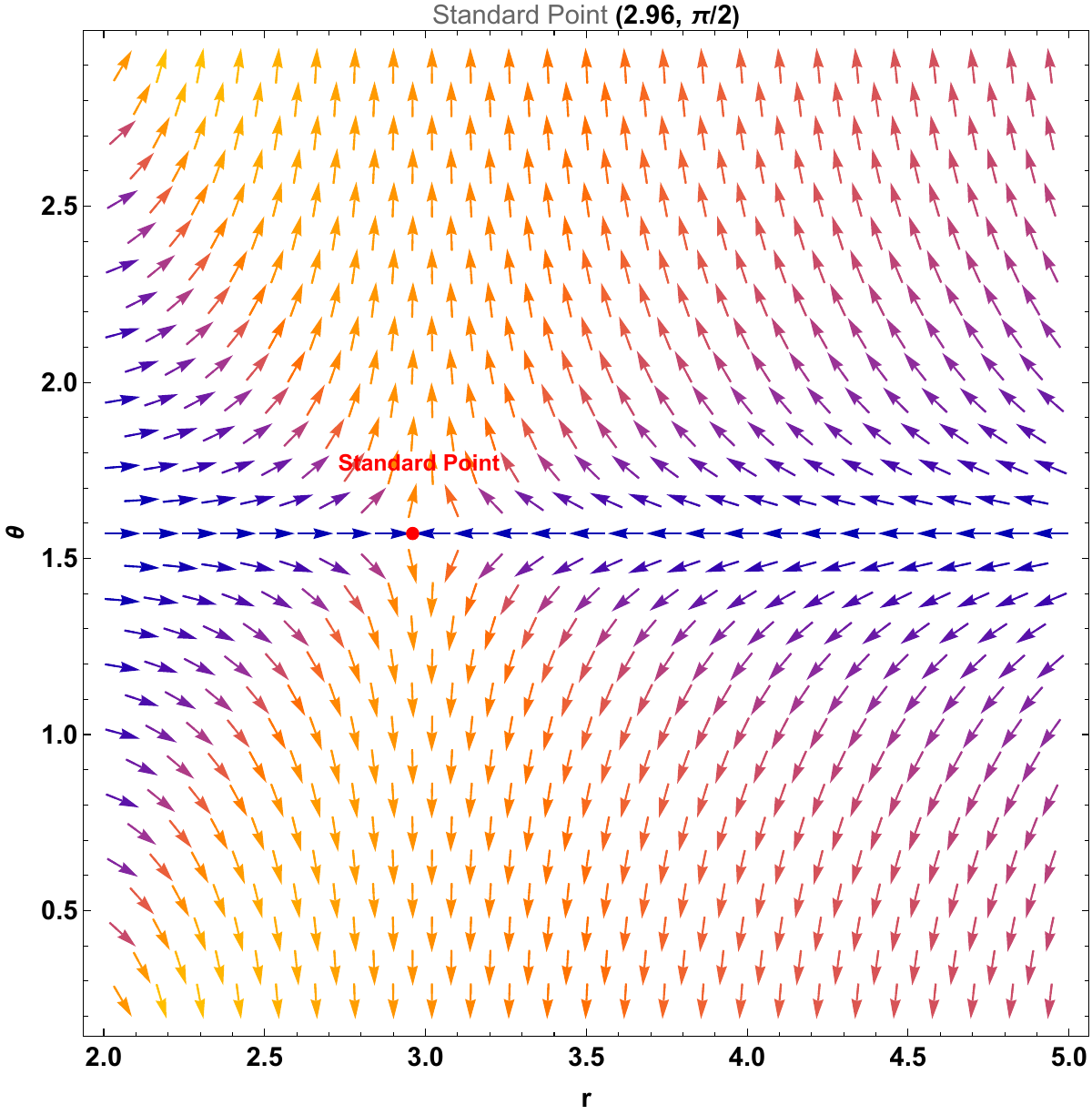}\\
    (a) $\ell=0,\,\alpha=0.1$ \hspace{6cm} (b) $\ell=0.1,\,\alpha=0$\\
    \caption{\footnotesize Comparison of the normalized vector representations. The arrows represent the normalized vector ${\bf n}$ on a portion of the $r-\theta$ plane for two BH solutions with $M=1$. The left panel corresponds to the Letelier BH solution, where the red dot is at $(r,\theta)=(r_\text{ph},\pi/2)$. The right panel depicts the holonomy-corrected Schwrazschild BH solution with the red dot is at $(r,\theta)=(r_0,\pi/2)$ with $r_0=r_\text{ph}-0.04$. The photon sphere radius $r_\text{ph}$ is listed in Table \ref{tab:1}. The left panel corresponding to the Letelier BH, while the right panel for holonomy-corrected Schwarzschild BH.}
    \label{fig:vector-comparison}
\end{figure}

In Figure~\ref{fig:vector-comparison}, we visualize the behavior of the normalized vector field $\mathbf{n}$ over a segment of the $r{-}\theta$ plane for two types of BH solutions: the Letelier and Holonomy-corrected Schwarzschild BHs.

In the limit $\ell \to 0$ corresponding to the absence of holonomy correction, the above result reduces as,
\begin{align}
    n_r&=-\frac{\left(1-\alpha-\frac{3\,M}{r}\right)}{\sqrt{(\left(1-\alpha-\frac{3\,M}{r}\right)^2+\left(1-\alpha-\frac{2\,M}{r}\right)\,\cot^2 \theta}},\label{cc11}\\
    n_{\theta}&=-\frac{\sqrt{1-\alpha-\frac{2\,M}{r}}\,\cot \theta}{\sqrt{(\left(1-\alpha-\frac{3\,M}{r}\right)^2+\left(1-\alpha-\frac{2\,M}{r}\right)\,\cot^2 \theta}}\label{cc12}.
\end{align}
Equations (\ref{cc11})--(\ref{cc12}) are the components of the unit vector field of topological photon rings in the background of Letelier BH solution.

Moreover, in the limit $\alpha=0$ corresponding to the absence of strings cloud, the normalized vector reduces as,
\begin{align}
    n_r&=-\frac{\sqrt{1-\frac{\ell}{r}}\,\left(1-\frac{3\,M}{r}\right)}{\sqrt{\left(1-\frac{\ell}{r}\right)\,(\left(1-\frac{3\,M}{r}\right)^2+\left(1-\frac{2\,M}{r}\right)\,\cot^2 \theta}},\label{cc13}\\
    n_{\theta}&=-\frac{\sqrt{1-\frac{2\,M}{r}}\,\cot \theta}{\sqrt{\left(1-\frac{\ell}{r}\right)\,(\left(1-\frac{3\,M}{r}\right)^2+\left(1-\frac{2\,M}{r}\right)\,\cot^2 \theta}}.\label{cc14}
\end{align}
Equations (\ref{cc13})--(\ref{cc14}) are the components of the unit vector field of topological photon rings in the background of holonomy corrected Schwarzschild BH solution.

We can also define the winding number, which is an important topological property, using the following formula,
\begin{equation}
    \omega_i=\frac{1}{2\pi} \oint_{C_i}\,d\Omega,\quad\quad \Omega=\frac{v_{\theta}}{v_r},\label{cc15}
\end{equation}
where $C_i$ denote a closed, smooth, and positively oriented curve that encapsulates solely the zero point of ${\bf v}$, while all other zero points lie external to it. The total topological charge associated with the PSs is then given by the sum of the winding numbers,
\begin{equation}
    Q=\sum_{i}\,\omega_i,\label{cc16}
\end{equation}
The presence of a zero point within a closed curve indicates that the total charge is equal to the winding number. Since every BH with a photon sphere likely possesses a topological charge at its photon sphere (PS), we can assign a distinct topological charge to each PS, which can be either \( +1 \) or \( -1 \). Furthermore, depending on the choice of the closed curve-which may encompass one or more zero points-the total topological charge enclosed by the curve may be \( +1 \), \( 0 \), or \( -1 \), as illustrated \cite{ref39}.

\section{Accretion disk radiation properties}\label{sec:5}

In this section, we analyze the properties of a thin accretion disk around the Holonomy corrected Schwarzschild BH in the presence of a cloud of strings. Following the Novikov–Thorne model, the accretion disk is assumed to be optically thick and geometrically thin near the BH. Under this assumption, the vertical entropy and pressure of the accreting matter are negligible, while efficient thermal radiation prevents heat buildup due to stresses and dynamical friction. The gas and dust in the disk move on nearly circular stable orbits, gradually shifting toward the inner edge of the accretion disk due to Reynolds stress generated by magnetic turbulence. This inward spiral toward the compact object results in strong electromagnetic radiation from the disk. 

The geometric and thermodynamic properties of the accretion disk are affected by the BH parameters, namely the cloud of strings parameter $\alpha$ and the LQG holonomy parameter $\ell$. In particular, these parameters influence the inner edge of the disk, the energy flux, and the resulting radiation spectrum.
The flux of electromagnetic radiation in the Novikov–Thorne framework is given by~\cite{Novikov:1973,Thorne:1974ve},
\begin{equation}
    F(r) = -\frac{\dot{M}_{0}}{4\pi \sqrt{g}} 
    \frac{\partial \Omega / \partial r}{(E - \Omega L)^{2}}
    \int_{r_{\rm ISCO}}^{r} (E - \Omega L)\,\frac{\partial L}{\partial r}\,dr,
    \label{eq:flux_nt}
\end{equation}
where $g$ is the determinant of the three-dimensional subspace ($g = \sqrt{-g_{tt} g_{rr} g_{\phi\phi}}$), $E$ and $L$ denote the specific energy and angular momentum of particles in circular orbits, $\Omega$ is the angular velocity, and $\dot{M}_{0}$ is the mass accretion rate. For simplicity, we normalize the flux by setting $\dot{M}_{0} = 1$, which is equivalent to analyzing the flux per unit accretion rate. Due to the complicated dependence on the spacetime geometry, Eq.~\eqref{eq:flux_nt} is computed numerically. 

\begin{figure}[ht!]
\centering
\includegraphics[width=0.4\textwidth]{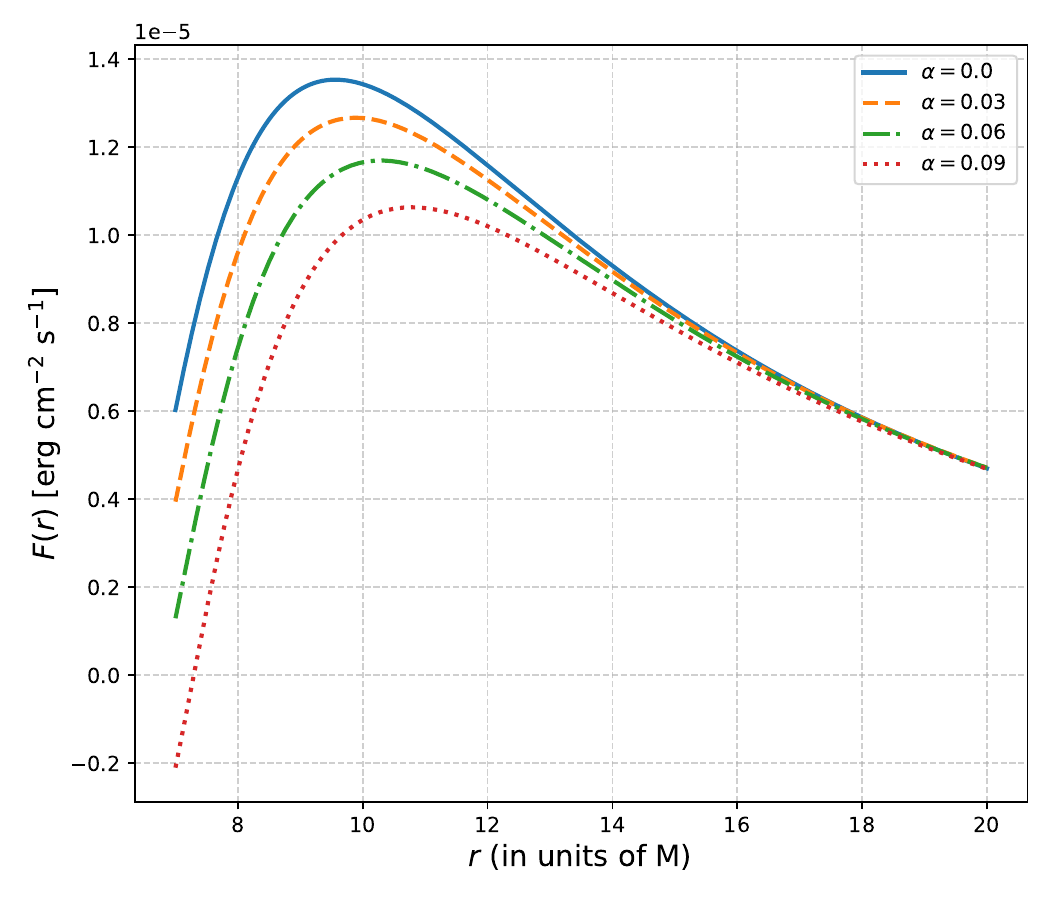}\quad
\includegraphics[width=0.4\textwidth]{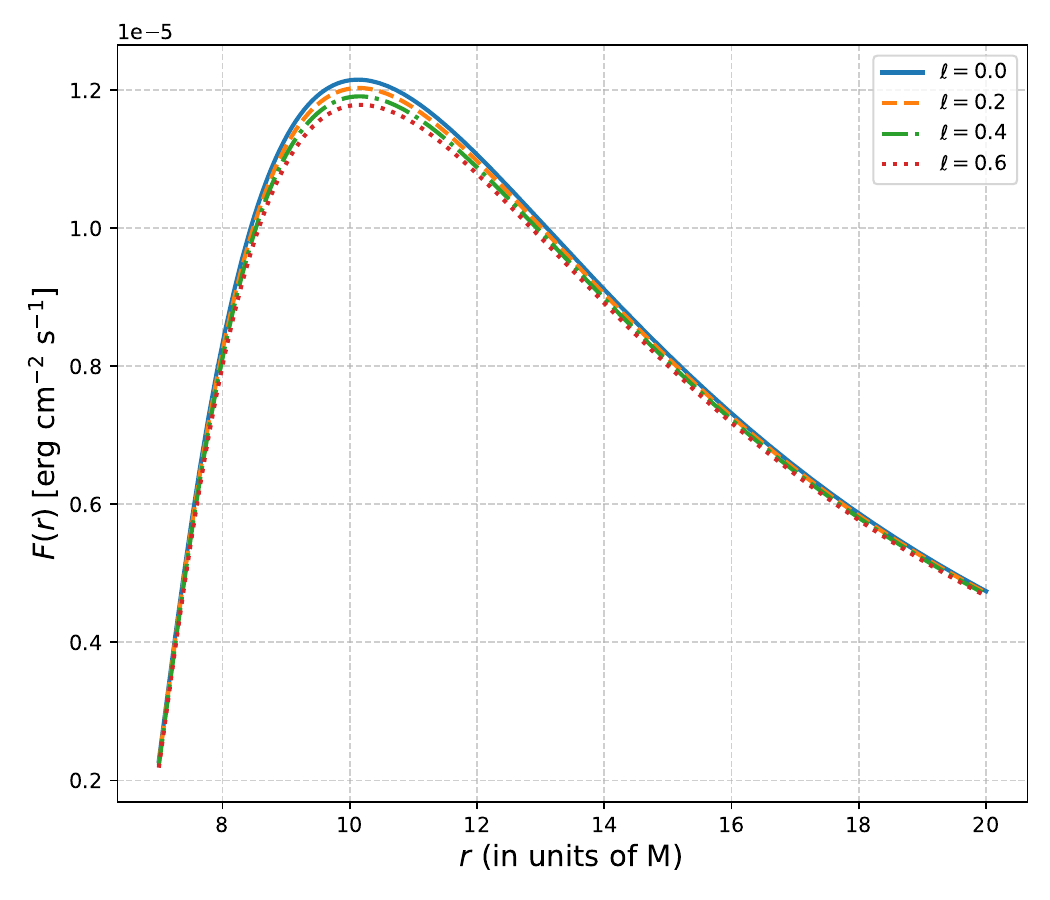}\\
(a) \hspace{6cm} (b)
\caption{\footnotesize The plot shows the radial dependence of the flux of electromagnetic radiation for different values of the $\alpha$ and $\ell$.} \label{fig:AD1}
\end{figure}

Furthermore, the flux of black-body radiation from the disk surface can be expressed as~\cite{Boshkayev:2023fft},
\begin{equation}
    F(r) = \sigma T^{4},
    \label{eq:flux_bb}
\end{equation}
where $\sigma$ denotes the Stefan–Boltzmann constant, and $T(r)$ is the local temperature of the disk. The radial dependence of the disk’s temperature can then be studied numerically for different values of the parameters $\alpha$ and $\ell$.

\begin{figure}[ht!]
\centering
\includegraphics[width=0.4\textwidth]{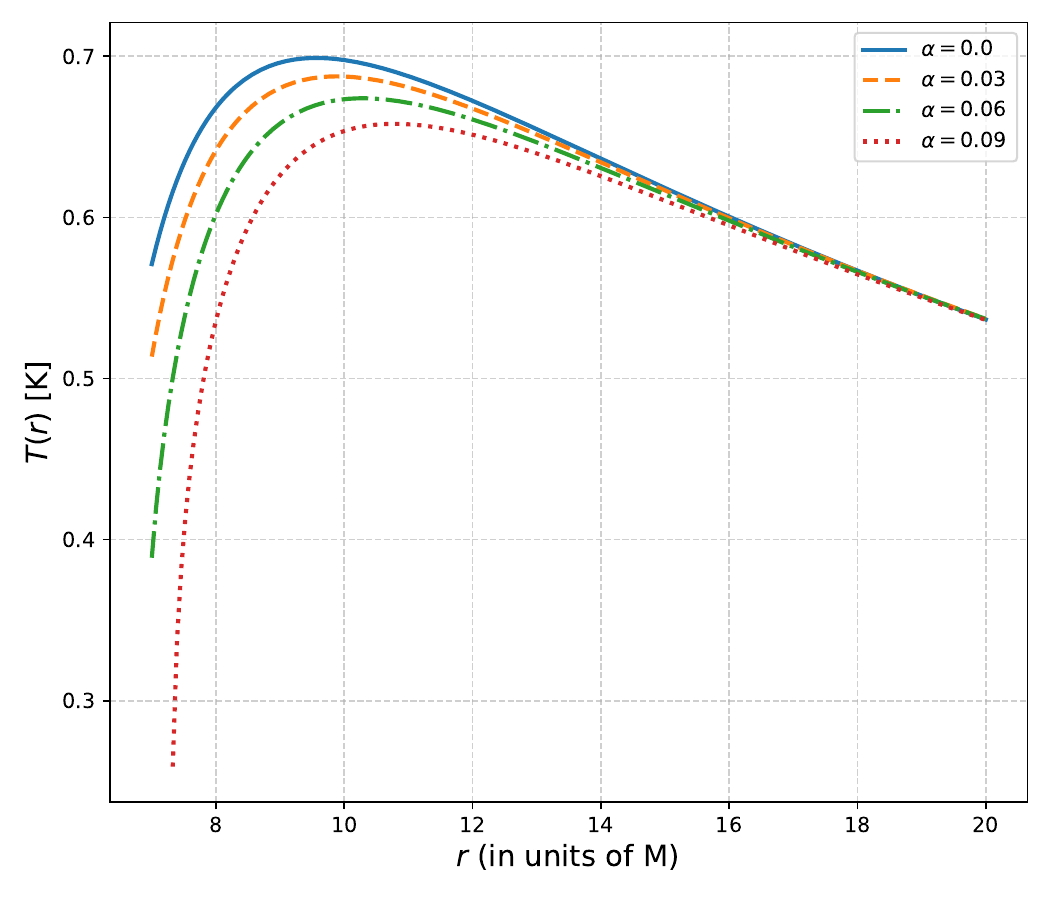}\quad
\includegraphics[width=0.4\textwidth]{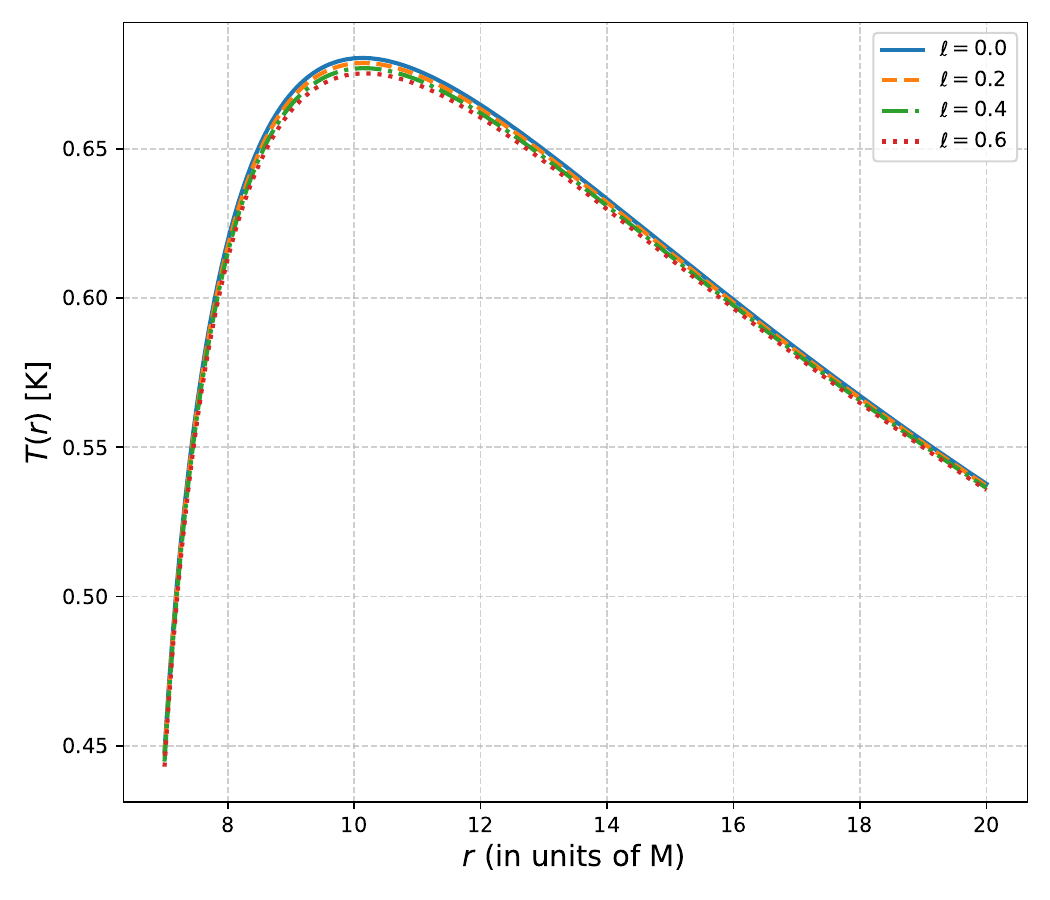}\\
(a) \hspace{6cm} (b)
\caption{\footnotesize The radial dependence of the disk temperature for different values of the $\alpha$ and $\ell$.} \label{fig:AD2}
\end{figure}

An important observable quantity is the differential luminosity of the disk, defined as~\cite{Boshkayev:2020kle},
\begin{equation}
    \frac{dL_{\infty}}{d \ln r} = 4 \pi r \sqrt{g}\, E\,F(r),
    \label{eq:diff_lum}
\end{equation}
which quantifies the energy output per logarithmic radial interval. The dependence of this quantity on the parameters $\alpha$ and $\ell$ provides further insights into the influence of the CoS and LQG corrections on the accretion disk structure. 

\begin{figure}[ht!]
\centering
\includegraphics[width=0.4\textwidth]{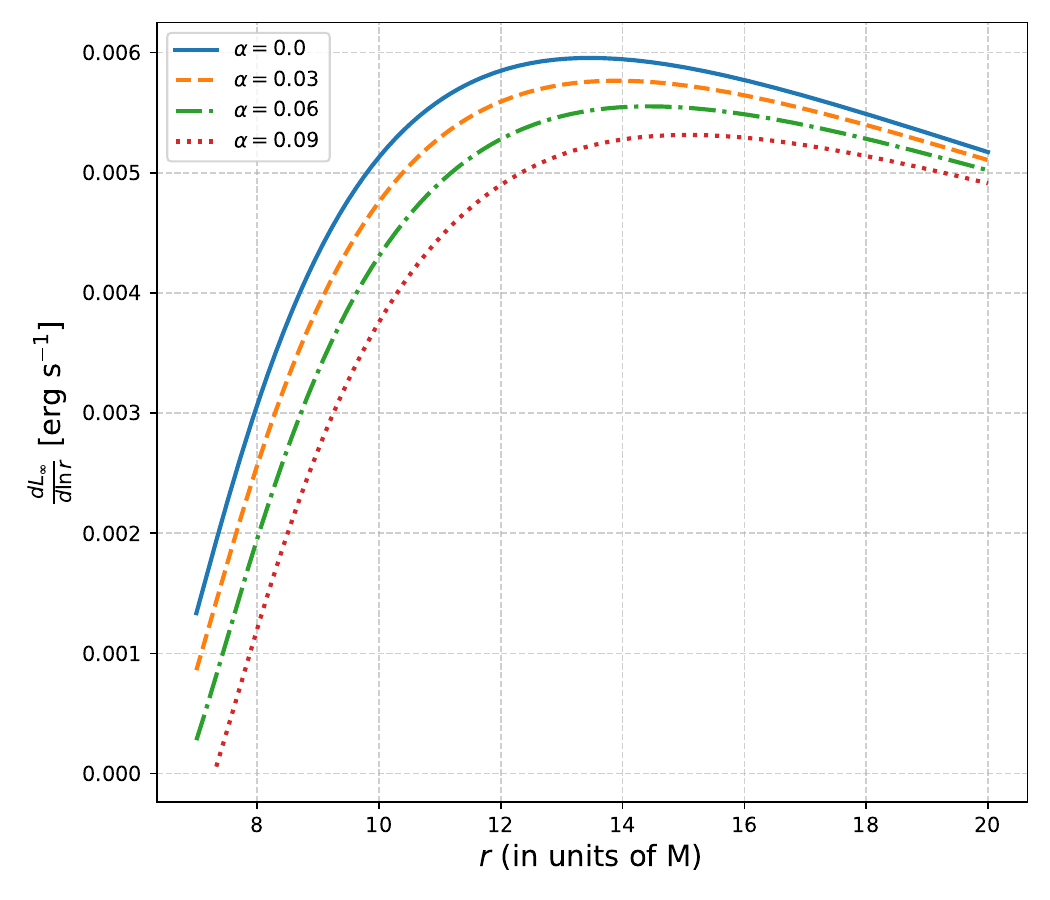}\qquad
\includegraphics[width=0.4\textwidth]{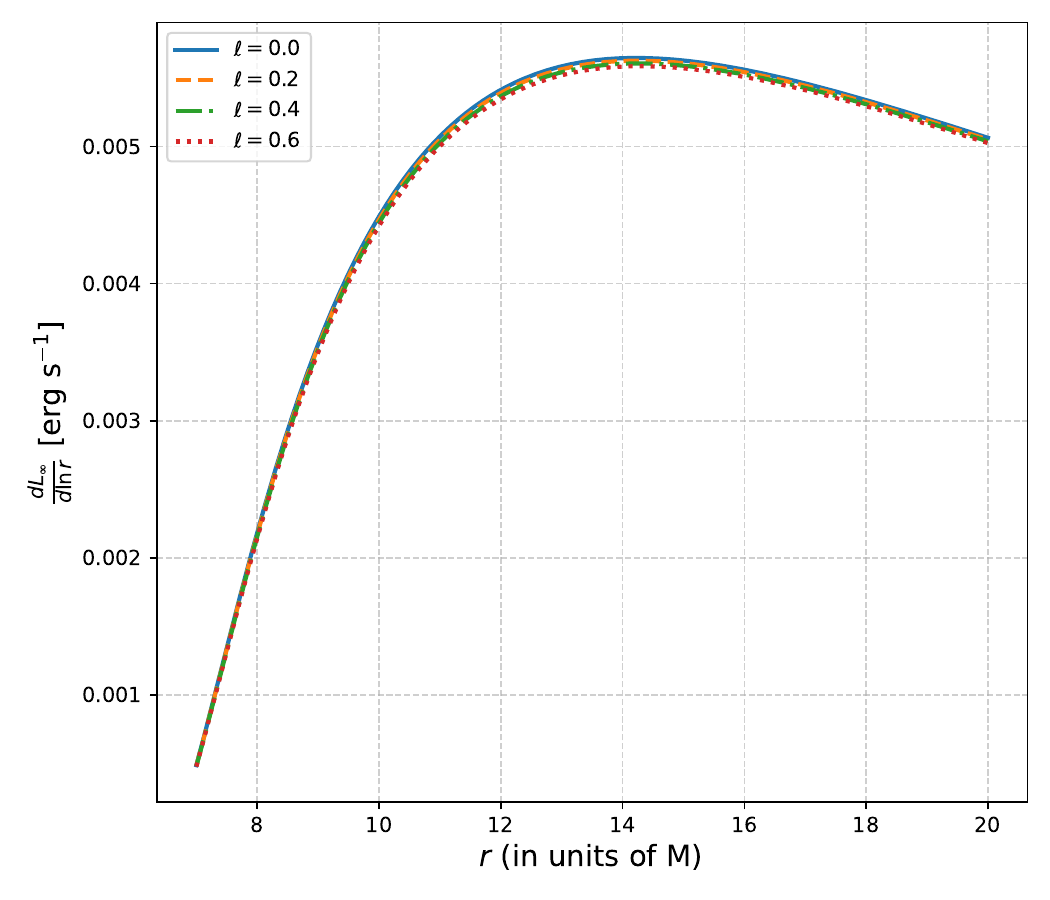}\\
(a) \hspace{6cm} (b)
\caption{\footnotesize The plot shows the radial dependence of the differential luminosity for different values of the $\alpha$ and $\ell$.} \label{fig:AD3}
\end{figure}

Finally, the total spectral luminosity of the disk can be modeled by assuming black-body emission at each radius. It takes the form~\cite{Torres:2002td,Guzman:2005bs},
\begin{equation}
    L(\nu) = \frac{16 \pi h \cos i}{c^{2}} \nu^{3}
    \int_{r_{\rm in}}^{r_{\rm out}} \frac{r\,dr}{\exp\left(\tfrac{h\nu}{k_{\rm B}T(r)}\right) - 1},
    \label{eq:spec_lum}
\end{equation}
where $i$ is the inclination angle of the disk, $h$ and $k_{\rm B}$ denote the Planck and Boltzmann constants, respectively, and $r_{\rm in}$ and $r_{\rm out}$ correspond to the inner and outer radii of the disk. In our analysis, we set $r_{\rm in} = r_{\rm ISCO}$ and $r_{\rm out} \rightarrow \infty$. This formulation allows us to investigate how the parameters $\alpha$ (cloud of strings) and $\ell$ (LQG corrections) modify the emission spectrum of the accretion disk around the HCSBH.

\begin{figure}[ht!]
\centering
\includegraphics[width=0.4\textwidth]{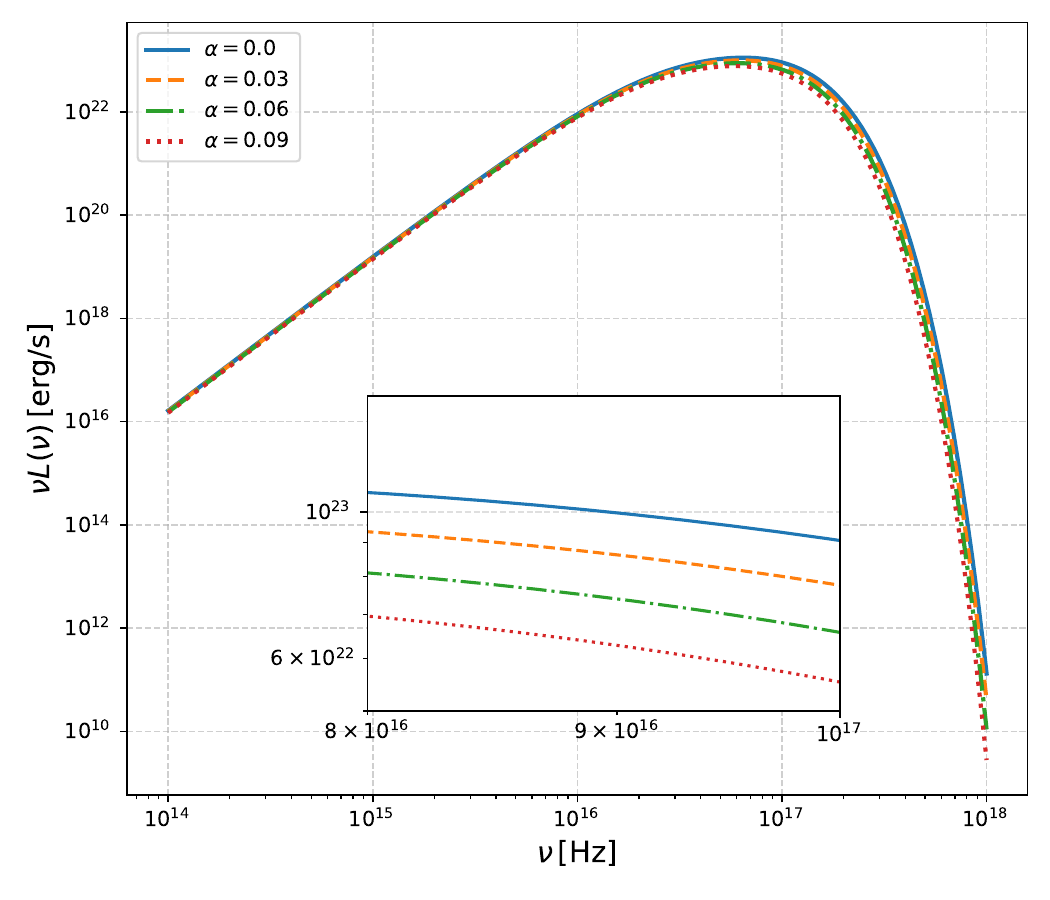}\qquad
\includegraphics[width=0.4\textwidth]{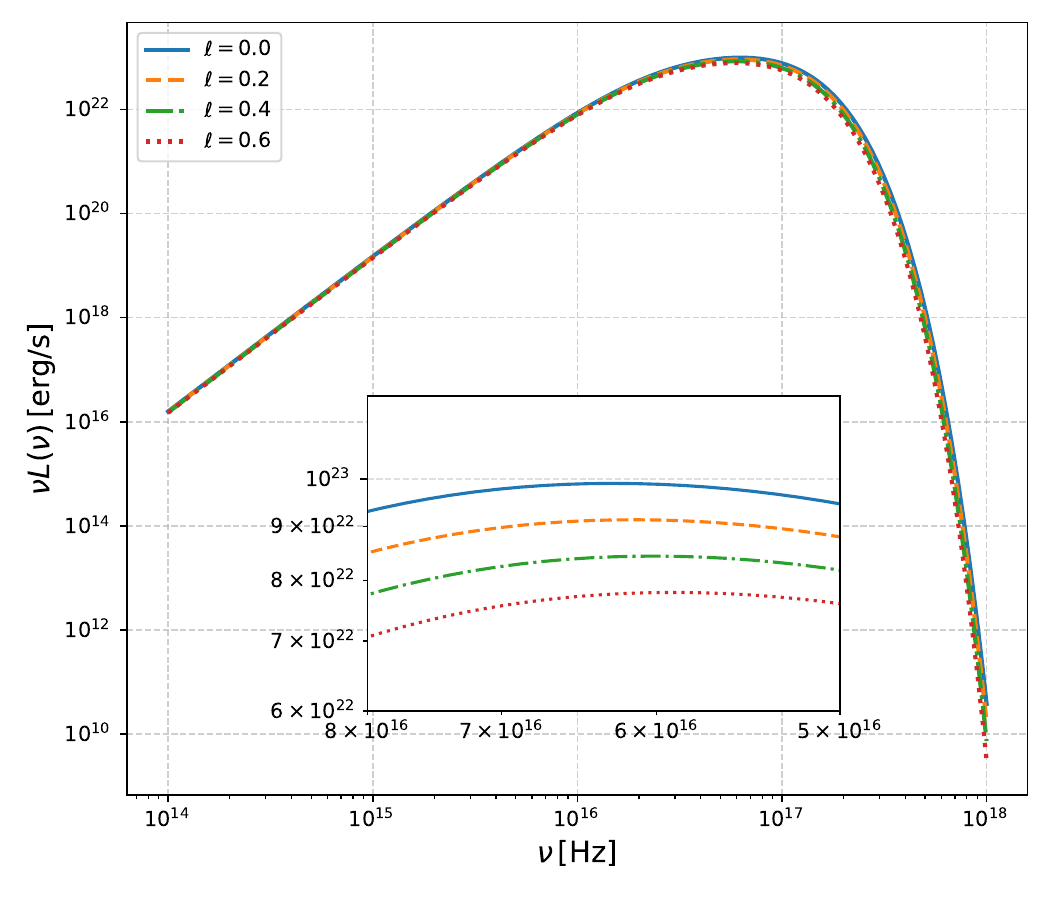}\\
(a) \hspace{6cm} (b)
\caption{\footnotesize The spectrum of the accretion disk for different values of the $\alpha$ and $\ell$. } \label{fig:AD4}
\end{figure}

Figures~\ref{fig:AD1}–\ref{fig:AD4} illustrate the effects of the cloud of strings parameter $\alpha$ and the LQG holonomy parameter $\ell$ on the observable quantities of the thin accretion disk around the holonomy corrected Schwarzschild BH. In Fig.~\ref{fig:AD1}, we present the radial dependence of the electromagnetic flux $F(r)$ emitted from the disk surface. As expected, the flux peaks near the inner edge of the disk and gradually decreases with increasing $r$. Both $\alpha$ and $\ell$ reduce the overall magnitude of the flux, indicating that the cloud of strings and LQG corrections suppress the efficiency of energy extraction from the accretion disk. The corresponding disk temperature profiles, obtained via the Stefan–Boltzmann law $F(r) = \sigma T^{4}$, are shown in Fig.~\ref{fig:AD2}. The temperature follows the same qualitative behavior as the flux, with a maximum near the innermost stable circular orbit (ISCO) and a monotonic decline at larger radii. Increasing either $\alpha$ or $\ell$ shifts the peak temperature downward, resulting in a cooler disk. This demonstrates that the spacetime modifications significantly affect the thermal structure of the accretion disk. In Fig.~\ref{fig:AD3}, we plot the differential luminosity $dL_{\infty}/d\ln r$, which measures the radial distribution of the energy output observed at infinity. The luminosity rises steeply near the ISCO, reaches a maximum in the inner disk region, and then decreases outward. The parameters $\alpha$ and $\ell$ reduce the amplitude of the luminosity profile, in agreement with the trends observed for $F(r)$ and $T(r)$. Finally, Fig.~\ref{fig:AD4} displays the disk spectra, represented by $\nu L(\nu)$ as a function of frequency. The spectra exhibit the typical multi-temperature blackbody shape, with a sharp rise in the low-frequency regime, a pronounced peak in the soft X-ray band, and an exponential cutoff at high frequencies. The peak luminosity decreases with larger $\alpha$ and $\ell$, and a slight shift of the peak frequency toward lower values is observed. These results confirm that the cloud of strings and LQG corrections imprint distinctive signatures on the accretion disk spectrum, which could in principle be probed through precise observations of BH accretion phenomena. 

\section{Conclusions}\label{sec:6}

Numerous studies had investigated the phenomenon of photon deflection in various curved spacetime backgrounds, including those generated by BHs (BHs), wormholes, and topological defect geometries. In the present study, we focused on the gravitational lensing properties, topological characteristics of light rings, and the lens equation in the background of a holonomy-corrected BH surrounded by a cloud of strings (CoS).

We obtained an exact expression for the bending angle in terms of elliptic integrals. Using this, we compared the Holonomy-corrected Schwarzschild black hole with other well-known black holes, showing how it reduces to standard cases in specific limits. We find that the inclusion of loop quantum gravity (LQG) and CoS parameters increases the deflection angle, and in the absence of the LQG parameter, varying the CoS parameter yields results in excellent agreement with Soares et al.\cite{HH4} (Fig.\ref{fig:defangle01}). Further, we analytically derived the deflection angle of photon trajectories in the weak-field limit and obtained explicit expressions revealing the dependence of light bending on the fundamental parameters of the BH geometry. It was found that the deflection angle was significantly influenced by the string cloud parameter $\alpha$ and the LQG holonomy correction parameter $\ell$. As shown in Figs.~\ref{fig:defangle1}--\ref{fig:defangle2}, the deflection angle decreased monotonically with increasing impact parameter. However, higher values of $\alpha$ and $\ell$ led to enhanced deflection, indicating an increase in the effective spacetime curvature due to the presence of string clouds and quantum gravity effects. Correspondingly, the total magnification of lensed images increased with both $\alpha$ and $\ell$, as illustrated in Figs.~\ref{fig:MAG1}--\ref{fig:MAG3}. This magnification effect was especially prominent when the source was positioned near the optical axis, emphasizing the sensitivity of gravitational lensing observables to the corrections introduced by CoS and LQG. These results suggested that lensing signatures could serve as viable observational probes of such modifications in the strong-field regime.

In addition to the analysis of lensing behavior, we studied the topological properties of the photon sphere associated with the holonomy-corrected BH surrounded by strings cloud. By constructing and examining a normalized vector field, we demonstrated that both $\alpha$ and $\ell$ significantly affected the photon ring topology. Figure~\ref{fig:vector} illustrated how variations in $\alpha$ modified the structure of the vector field in the equatorial plane at the standard point $(r, \pi/2)$. Furthermore, Fig.~\ref{fig:vector-comparison} presented a comparison between the Letelier BH solution and the holonomy-corrected Schwarzschild solution. This comparison highlighted the distinct imprints of the parameters $\alpha$ and $\ell$ on the normalized vector field, offering additional insights into the geometrical and topological effects induced by CoS and LQG corrections.

Finally, we investigated the influence of these parameters on the physical properties of a thin accretion disk surrounding the BH. As shown in Figs.~\ref{fig:AD1}--\ref{fig:AD4}, increasing values of $\alpha$ and $\ell$ led to a suppression of key accretion disk quantities, including the flux profile $F(r)$, the temperature distribution $T(r)$, and the differential luminosity $dL_{\infty}/d\ln r$. This suppression reflected a decrease in the efficiency of energy extraction from the accreting matter. The corresponding spectral energy distribution also exhibited a reduction in the peak luminosity and a slight shift of the peak frequency toward lower values, indicating a cooler disk profile in the presence of string clouds and LQG modifications.

In conclusion, our findings have shown that the presence of string clouds and holonomy corrections left measurable and distinct signatures on both gravitational lensing and accretion disk observables. These effects offered potential avenues for placing observational constraints on quantum gravity-inspired modifications in strong gravitational fields.

For future research, we propose several possible extensions of the present work. A natural direction would be to incorporate rotation into the BH metric in conjunction with the parameters $\alpha$ and $\ell$, thereby providing a more generalized and realistic framework for studying astrophysical BHs. In addition, thermodynamics properties including the Joule-Thomson expansion \cite{GMM2,GMM3,GMM4,GMM5,GMM6} and thermodynamic topology \cite{TP1,TP2} of the BH system could provide deeper insights into the interplay between CoS effects and holonomy-corrections in BH thermodynamics. Extending the analysis to perturbations of scalar, vector and fermionic field, alongside the computation of greybody factors and scattering cross-sections, would further enrich our understanding of quantum processes in these exotic space-times.

{\small
\section*{Acknowledgments}

F.A. acknowledges the Inter University Centre for Astronomy and Astrophysics (IUCAA), Pune, India for granting visiting associateship. The author, SK, sincerely acknowledges IMSc for providing exceptional research facilities and a conducive environment that facilitated his work as an Institute Postdoctoral Fellow.

}

\end{document}